\documentclass[twocolumn]{aastex631}
\usepackage{amsfonts,amsmath,appendix,graphics,graphicx,color,hyperref}

\shorttitle{GRB Host Galaxies at $z\sim5$}
\shortauthors{Sears et al.}

\begin{document}

\title{Constraints on the $z\sim5$ Star-Forming Galaxy Luminosity Function From \textit{Hubble Space Telescope} Imaging of an Unbiased and Complete Sample of Long Gamma-ray Burst Host Galaxies}

\author[0000-0001-8023-4912]{Huei Sears}
\affiliation{Center for Interdisciplinary Exploration and Research in Astrophysics (CIERA), Northwestern University, Evanston, IL 60202, USA}
\affiliation{Department of Physics and Astronomy, Northwestern University, Evanston, IL 60208, USA}

\author[0000-0002-7706-5668]{Ryan Chornock}
\affiliation{Department of Astronomy, University of California, Berkeley, CA 94720-3411, USA}

\author[0000-0002-1468-9668]{Jay Strader}
\affiliation{Center for Data Intensive and Time Domain Astronomy, Department of Physics and Astronomy, Michigan State University, East
Lansing MI, USA}

\author[0000-0001-8472-1996]{Daniel A. Perley}
\affiliation{Astrophysics Research Institute, Liverpool John Moores University, Liverpool Science Park, 146 Brownlow Hill, Liverpool L3 5RF, UK}

\author[0000-0003-0526-2248]{Peter K. Blanchard}
\affiliation{Center for Interdisciplinary Exploration and Research in Astrophysics (CIERA), Northwestern University, Evanston, IL 60202, USA}
\affiliation{Department of Physics and Astronomy, Northwestern University, Evanston, IL 60208, USA}

\author[0000-0003-4768-7586]{Raffaella Margutti}
\affiliation{Department of Astronomy, University of California, Berkeley, CA 94720-3411, USA}
\affiliation{Department of Physics, University of California, 366 Physics North MC 7300, Berkeley, CA 94720, USA}

\author[0000-0003-3274-6336]{Nial R. Tanvir}
\affiliation{School of Physics and Astronomy, University of Leicester, University Road, Leicester, LE1 7RH, UK}

\begin{abstract}

We present rest-frame UV \textit{Hubble Space Telescope} imaging of the largest and most complete sample of 23 long duration gamma-ray burst (GRB) host galaxies between redshifts 4 and 6.  Of these 23, we present new WFC3/F110W imaging for 19 of the hosts, which we combine with archival WFC3/F110W and WFC3/F140W imaging for the remaining four.  We use the photometry of the host galaxies from this sample to characterize both the rest-frame UV luminosity function (LF) and the size-luminosity relation of the sample.  We find that when assuming the standard Schechter-function parameterization for the UV LF, the GRB host sample is best fit with $\alpha = -1.30^{+0.30}_{-0.25}$ and  $M_* = -20.33^{+0.44}_{-0.54}$ mag, which is consistent with results based on $z\sim5$ Lyman-break galaxies.  We find that $\sim68\%$ of our size-luminosity measurements fall within or below the same relation for Lyman-break galaxies at $z\sim4$.  This study observationally confirms expectations that at $z\sim5$ Lyman-break and GRB host galaxies should trace the same population and demonstrates the utility of GRBs as probes of hidden star-formation in the high-redshift universe.  Under the assumption that GRBs unbiasedly trace star formation at this redshift, our non-detection fraction of 7/23 is consistent at the $95\%$-confidence level with $13 - 53\%$ of star formation at redshift $z\sim5$ occurring in galaxies fainter than our detection limit of $M_{1600\mathrm{\AA}} \approx -18.3$ mag.  

\end{abstract}


\section{Introduction} \label{sec:intro}

Long duration gamma-ray bursts (GRBs) have been theoretically \citep{Paczynski1986, Woosley1993} and observationally associated with the deaths of massive stars and specifically with Type Ib/c-BL supernovae (SNe).  These SNe result from the core collapse of 
a progenitor star that has completely lost its hydrogen shell and most-to-all of its helium shell, with the ``BL" designation in reference to the fast moving SN ejecta resulting in broad-lined (BL) emission features (\citealt{Galama1998, MacFadyen1999, Hjorth2003, WoosleyBloom2006}, and \citealt{HjorthBloom2012, Cano2017} for recent review).  There are two main observational components to a GRB - the initial gamma-ray prompt emission believed to be from dissipation processes within the GRB jet and the multi-wavelength afterglow powered by the synchrotron emission originating from the jet's deceleration into the local environment \citep{Chavalier1999, Miceli2022}.

GRB follow-up and afterglow studies were revolutionized with the launch of the \textit{Neil Gehrels Swift Observatory} (\textit{Swift}; \citealt{Gehrels2004}). The X-ray telescope on board (XRT; \citealt{Burrows2005}) has the ability to localize the GRB afterglow to few arcsecond precision allowing for ground based observations.  As long-duration GRBs are known to predominantly occur within the half-light radius and within the bright, star-forming regions of their host galaxies \citep{Fruchter2006, Svensson2010, Blanchard2016, Lyman2017}, this precise afterglow-enabled localization often allows for robust host identification. The extreme luminosity ($\sim10^{53}$ erg s$^{-1}$) of the GRB makes them observable to cosmological distances, with the currently most distant GRB 090429B photometrically estimated to have $z=9.4$ \citep{grb090429B_redshift}.

High-redshift ($z>3$) star-forming galaxies are primarily identified using the Lyman break technique in which the wavelength of the Lyman break is determined via photometric dropout \citep{Steidel1996}. Studies of star-forming galaxies benefit from large-number statistics and deep observations and, prior to \textit{JWST}, extend through $z\sim9$ (see e.g., \citealt{Stark_review} for a recent review). Surveys from \textit{JWST}, including early data release and dedicated programs like The Cosmic Evolution Early Release Science Survey (CEERS; \citealt{CEERS1}), The GLASS JWST Early Release Science Program (GLASS-JWST; \citealt{GLASS}), and the JWST Advanced Deep Extragalactic Survey (JADES; \citealt{JADES_JWST}) have allowed for analysis of these galaxies to continue to even greater redshift ($z\sim13$).  An important characterization of Lyman-break galaxies is the UV luminosity function. This function is a fit to a histogram of these galaxies and allows for an estimate of the percentage of undetectable star-formation through extrapolation of the fit to faint magnitudes.  It is well defined at the bright end ($M_{UV}<-15$ mag) \citep{Finkelstein2015, Bouwens_2021a,Bouwens2022,Harikane2023a, Harikane2023b, CEERS1} with the generally assumed \cite{schechterfunction} being fit to measurements from thousands of galaxies.

Observations of Lyman-break galaxies, however, only offer a view of the star formation that can be directly observed and are therefore implicitly biased against faint galaxies.  Since the ability to detect a GRB is independent of the luminosity of its host galaxy, and the detection of a GRB implies the existence of a galaxy at that location, GRBs offer a way to characterize faint and otherwise unobserved star formation, such as that which is dust-obscured or intrinsically faint.  Constraining the amount of star formation that would otherwise go undetected, especially at high redshift, is key for determining how large a role this star-formation played in reionizing the universe. 

In the low-redshift Universe ($z<2$) GRB host galaxies have been shown to have smaller sizes, lower masses and lower metallicities than the general star-forming galaxy population \citep{Stanek2006, Kewley2007, Levesque2010, Svensson2010, Han2010, GrahamFruchter2013, Perley2013, Palmerio2019}. These biases are thought to be a consequence of the preference for a GRB progenitor to form and explode in low-metallicity environments, with low-metallicity star-forming galaxies being smaller and less massive than the general sample \citep{Mannucci2010, Palmerio2019}.  The nature of this preference, both physical and functional, is still actively debated: some studies have theorized multiple metallicity-dependent paths for GRB creation \citep{Trenti2015}, while some have found evidence for a host-galaxy stellar metallicity threshold above which GRBs are rare (i.e., it allows for the possibility of a pocket of lower-Z star-formation within a high-Z galaxy). Below this threshold, GRBs seem to trace star formation in an unbiased way (though there is uncertainty on the value of this threshold ($Z<Z_{\odot}$: \citealt{SHOALS_II}; $Z<0.7Z_{\odot}$: \citealt{Palmerio2019}).

The bias of the GRBs in host galaxy mass and size is consistent with being largely a by-product of the metal-aversion \citep{SHOALS_II}, and so, as the average metallicity of the Universe decreases with increasing redshift, the differences in the characteristics of GRB host galaxies as compared to those of actively star-forming galaxies should decrease toward triviality.  Indeed, up to $z\sim4$, comparisons of the two galaxy samples have followed this expectation when characterized by the mass-metallicity relation \citep{Levesque2010, Laskar2011, GrahamSchadyFruchter2019}, the UV luminosity function \citep{Greiner2015,TOUGH}, and in direct size and stellar mass measurements \citep{Schneider2022}.  Comparisons at higher redshift ($z\sim6$) also support these results but are significantly limited in precision due to the small number of localized GRBs with confirmed redshifts at this redshift range \citep{Tanvir2012,130606A_McGuire}.

In this work, we present new \textit{Hubble Space Telescope (HST)} observations of the largest complete sample of GRB host galaxies at $z\sim5$, to significantly improve these comparisons at the highest possible redshifts with currently available data. In Section \ref{sec:obs}, we describe our observations and host identification methods.  We present our formalism, modeling, and analysis of the UV luminosity function and size-luminosity relation of the GRB host sample and compare to that of Lyman-break galaxies in Section \ref{sec:discussion}.  We conclude with presentation and discussion of our non-detection fraction and its implications toward the amount of undetectable star formation.  We use a cosmological model with $H_0 = 70\ \text{km\ s}^{-1}\ \text{Mpc}^{-1}$, $\Omega_0 = 0.3$, and $\Omega_{\Lambda} = 0.7.$  Uncertainties are reported as the Gaussian-equivalent one-sigma, unless otherwise stated.
 
\section{Observations}\label{sec:obs}

\subsection{Sample Selection}
We define selection criteria for our $z\sim5$ GRB host galaxy sample to minimize selection bias while maximixing completeness.  Our initial selection criteria were:

\begin{itemize}
    \item The GRB has a spectroscopic or photometric redshift of $4<z<6$.
    \item Deep observations at the GRB location were performed with the \textit{Spitzer Space Telescope} (\textit{Spitzer}; \citealt{Werner2004}).
    \item The GRB was detected with \textit{Swift} prior to mid-2015 (the date is a by-product of the \textit{Spitzer} requirement) and has a localization $\leq2''$.
    \item The line of sight along the GRB direction has low Galactic extinction, $E(B-V) < 0.2$ mag.
\end{itemize}

From this first-round sample, we use the following criteria to determine the final sample:

\begin{enumerate}
    
    \item The GRB was included in one of the four following uniform samples: The Optically Unbiased GRB Host Survey (TOUGH; \citealt{TOUGH}), A Complete Sample of Bright Swift Long Gamma-Ray Bursts (BAT6; \citealt{Salvaterra2012}), or The X-shooter GRB afterglow legacy sample (XS-GRB; \citealt{XSGRB}), The Swift GRB Host Galaxy Legacy Survey (SHOALS; \citealt{SHOALS_1}), or otherwise met the criteria to be included in the SHOALS sample but occurred outside of the project timeline, \textbf{OR}

    \item The GRB was rapidly observed with a NIR imager on a $>1$m telescope, such as the Palomar 60-inch Telescope (P60; \citealt{Cenko2006}), the Peters Automated Infrared Imaging Telescope (PAIRITEL; \citealt{Bloom2006PAIRITEL}), or the Gamma-Ray Burst Optical/Near-Infrared Detector (GROND; \citealt{Greiner2008}) on the MPG/ESO 2.2-meter telescope.

\end{enumerate}

From these criteria, we populate a sample of 19 GRBs for host galaxy follow-up.  We add to this a randomized subsample of four GRBs (050505, 060223A, 140304A, and 140311A) of the 10 events which did not pass the final sample criteria.  After investigating the selection criteria for each of the uniform samples, these four GRBs had been excluded due to a small Sun hour angle separation, too high of a declination, were not observed with XRT within 10 minutes of the \textit{Swift} Burst Alert Telescope (BAT; \citealt{SwiftBAT}) trigger, or had too low a fluence ($S_{15-150 keV}$).  These properties, as well as the non-existence of rapid NIR follow-up, have no dependence on the characteristics of the GRB host galaxy and the inclusion of these four GRBs has no effect on the uniformity of our GRB host galaxy sample.

\subsection{Hubble Space Telescope Imaging}
We present new \textit{HST}/WFC3-IR imaging for 19 galaxies in our sample (ID: 15644, PI: Perley), while the remaining four had archival imaging available, which we detail in the following section.  The 19 host galaxies from our program were imaged using the F110W filter: galaxies with redshift $z < 4.8$ were observed over two orbits (average exposure time, 4900 s), while those with $z > 4.8$ were observed over three orbits (average exposure time, 7400 s).  Across our redshift range, the central rest-frame wavelength of F110W converts to 1650--2260~\AA, which samples the rest-frame UV emission. 

We use archival imaging for four sources which were previously observed by \textit{HST}.  The host galaxies of GRBs 060223, 060522, and 060927 were also imaged using WFC3/F110W (ID: 11734, PI: Levan) with 3 orbits for the fields of 060223 and 060522 and 5 orbits for the field of 060927.  The host of GRB 130606A was imaged using WFC3/F140W (ID: 13831, PI: Tanvir) over 4 orbits.  At a redshift of z = 5.913 (\cite{130606A_MMT_redshift}), the central wavelength of F140W translates to 2014~\AA, which is comparable to the observations of the other objects in the sample.
 
The reduced (i.e., flat fielded, charge transfer efficiency (CTE) corrected, dark subtracted) and ICRS-aligned \textit{HST} images were downloaded from the Barbara A. Milkulski Archive for Space Telescopes (MAST) (see chapters 2 and 3 of \citealt{wfc3datahandbook} for details on this reduction).  To drizzle the \textit{HST} frames and achieve a resolution past the instrument limitation, we use \texttt{Astrodrizzle} \citep{AstroDrizzle} with \texttt{final\_pixfrac} = 0.8 and \texttt{final\_scale} = 0.065 for consistency with previous GRB host galaxy \textit{HST} analysis (e.g., \citealt{Blanchard2016}). 

\subsection{Afterglow Localizations}

Our analysis requires robust and accurate GRB localizations in order to identify the host galaxy of each GRB, and for that purpose, when possible, we use imaging of the optical afterglow. We were able to use optical afterglow imaging for all but three sources in our sample.  For these three sources with no optical/NIR afterglow imaging available, we use their position as reported from \textit{Swift}-XRT (GRBs 050803 and 050922B; \citealt{GoadXRT}) or from the Karl G. Jansky Very Large Array (VLA; \citealt{RPerley2011}) (GRB 140304A; \citealt{140304A_VLA}). 

Optical afterglow images were collected from the public archives of the Low Resolution Imaging Spectrometer at the W.M. Keck Observatory (Keck-LRIS; \citealt{Oke1995}), the Gemini-North/South Multi-Object Spectrograph at the Gemini North/South Observatory (GMOS-N/S; \citealt{Hook2004}), the Palomar 60-inch Telescope at Palomar Observatory (P60; \citealt{Cenko2006}), Very Large Telescope (VLT), the Rapid Eye Mount Telescope at La Silla Observatory (REM\footnote{\url{https://www.eso.org/public/teles-instr/lasilla/rem/}}), the Device Optimized for the LOw RESolution (DOLORES, in short LRS\footnote{\url{https://www.tng.iac.es/instruments/lrs/}}) at the Telescopio Nazionale Galileo (TNG), and the Ultraviolet/Optical Telescope onboard \textit{Swift} (UVOT; \citealt{Roming2005}).  To reduce images from Keck-LRIS, we use the \texttt{LPIPE} pipeline \citep{LPIPE}.  When possible, we use the reduction pipelines embedded within the archive services. We otherwise use standard reduction steps such as flat division, bias subtraction, and image stacking.  Centroid positions for each afterglow were measured using \texttt{Source Extractor} \citep{SourceExtractor}.  Imaging and reduction steps for each GRB afterglow are detailed in the Appendix with additional references in Table \ref{tab:afterglow_info}. 

\subsection{Astrometric Alignment}
Many of the afterglow images had an initial world coordinate system (WCS) assigned by the data archive.  For those that did not, we upload the afterglow image to Astrometry.net \citep{AstrometryNet} to get a preliminary WCS assignment.  To align the afterglow images to the \textit{HST} images, we used \texttt{TweakReg} \citep{AstroDrizzle}.  In the first alignment attempt, we use a catalogue of \textit{Gaia} sources within $2'$ of the afterglow position.  If this failed or if there were fewer than 6 catalogue sources in the \textit{HST} image, we instead used a catalogue of at least 6, but often $>10$, matching sources (all of the bright and unsaturated stars and sometimes bright galaxies) between each afterglow and \textit{HST} image pair.  These sources were selected from visual inspection in \texttt{SAOImageDS9} (DS9; \citealt{DS9_cite}).  The alignment was deemed successful when common sources were aligned to within approximately 1 \textit{HST} pixel = 0.065$''$.   In the case of GRB 060223, there was only one source (a saturated star) in common between the two images, and so we instead aligned each image separately to the \textit{Gaia} DR2 catalogue.  Details on alignment steps for each source are in the Appendix (\ref{sec:appendix}).

The afterglow positions found by \texttt{Source Extractor} were then converted from pixel to WCS coordinates for use in host galaxy identification in the corresponding \textit{HST} image.  To quantify the uncertainty on the position of the afterglow, reported in Table \ref{tab:afterglow_info}, we add in quadrature the uncertainty in the afterglow centroid from \texttt{Source Extractor} and the root mean square (RMS) of the astrometric match to the \textit{HST} image of the host.  When optical afterglow imaging was unavailable, we list the uncertainty reported in the literature (GRB 140304A; \citealt{140304A_VLA}) or the \textit{Swift}-XRT catalogue (GRBs 050803 and 050922B; \citealt{GoadXRT}). 
All but two afterglows (GRBs 050803 and 050922B, for which only \textit{Swift}-XRT imaging was available) were localized to better than $0.5''$, with a median localization uncertainty of $\sim0.06''$.

For all but one case (GRB\ 050922B), if there was a galaxy coincident within the afterglow uncertainty region, we designate that as the host of the GRB, as lower-redshift GRB afterglows are found near the centers of their host galaxy \citep{Blanchard2016}.  Within the afterglow uncertainty region of GRB\ 050922B, there are two galaxies: a compact source and a merging system.  In agreement with \cite{SHOALS_1}, we designate the merging system as the host of this GRB.  The identification of this galaxy as the host is elaborated upon in the next section.  If there was no galaxy within the region, we classified this as a non-detection for the host galaxy.  Details on the detection classification for each host are in the Appendix, and excerpts of the \textit{HST} imaging with afterglow positions, their three-sigma uncertainty regions, and host galaxy identifications are shown in Figure \ref{fig:Stamp_Collage}.

\subsection{$P_{cc}$\ Calculations}
We consider the false alarm probability for our claimed host galaxies and nearby sources to our claimed non-detections.  The false alarm probability is the chance of an unrelated galaxy being within the measured proximity to the line of sight to the GRB.  When the afterglow is well-localized, this probability is largely dependent on the offset from the afterglow and the apparent magnitude of the putative host.  We calculate the probability of chance coincidence ($P_{cc}$) following methods in \cite{Bloom2002} and using $P_{cc} =  1 - e^{-\pi\times R_{e}^2\times \sigma(\leq m)}.$  $R_{e}$ is taken to be the maximum of \Big[$3\sqrt{\sigma_{TIE}^2+\sigma_{AG}^2}$,$\sqrt{R^2+4\times R_{eff}^2}$\Big], where $\sigma_{TIE}$ is the uncertainty in the astrometric tie between the afterglow and galaxy positions, and $\sigma_{AG}$ is the uncertainty in the afterglow position. $R$ is the offset of the considered galaxy from the center of the afterglow, and $R_{eff}$ is the half-light radius of this considered galaxy. $\sigma(\leq m)$ is calculated from summing galaxy number densities below the measured $m_{F110W}$ in Tables 3 and 4 in \cite{Metcalfe2006}.  

For our detections, we calculate the $P_{cc}$ for the putative host.  Only four of the 16 putative host galaxies had $P_{cc} > 0.1$.  These were the host galaxies of GRBs 050803 ($P_{cc} = 0.98$), 050922B ($P_{cc} = 0.99$), 071025 ($P_{cc} = 0.12$), and 140614A ($P_{cc} = 0.21$).  These four cases include our two GRBs with only \textit{Swift}-XRT positions available (GRBs 050803 and 050922B) and two sources with the next largest afterglow positional uncertainty.  In the cases of 050803 and 050922B, these GRB were included in our sample due to the photometric redshifts of the claimed host galaxies \citep{SHOALS_1}, and so we continue analysis with the assumption that these are the host galaxies of these GRBs.  We repeated our analysis in Section 3 treating these hosts as non-detections, and found that the best-fit LF parameters are consistent to within 1 sigma, so our results are not strongly sensitive to the uncertainty in these host associations.  In the other two cases, these were the only sources within the afterglow uncertainty region, and so we classify them as the host galaxy of their respective GRB.  Details on each $P_{cc}$ are in the Appendix.

For our non-detections, we calculate the $P_{cc}$ for all sources detected by \texttt{Source Extractor} within a $5'' \times 5''$ box centered on the afterglow position reported in Table \ref{tab:afterglow_info}.  Only two of the 21 nearby sources in the $5''$ fields of our non-detections had $P_{cc}<0.1$.  These two sources (one each in the fields of GRBs 060927 and 100219A) were confirmed to have a lower redshift than each respective GRB and are therefore not the host galaxies.  The galaxy in the field of GRB 060927 was detected in \textit{R}-band VLT imaging \citep{Basa2012} and has a redshift $z<4$, which is incompatible with the spectroscopic afterglow redshift of $z=5.467$ reported in \cite{Ruiz-Velasco2007}.  The galaxy in the field of GRB 100219A was spectroscopically confirmed to have $z = 0.217$ in \cite{100219A_GMOS_S}, which is incompatible with the spectroscopic afterglow redshift of $z = 4.667$ for GRB 100219A \citep{XSGRB}.  Because all other detected candidate host galaxies have $P_{cc} > 0.1$, we report the host galaxies of these seven GRBs as non-detections.  Details on each $P_{cc}$ are in the Appendix (\ref{sec:appendix}).

\subsection{HST Photometry}

We measure apparent magnitudes of all detected GRB host galaxies with \texttt{Source Extractor} using \texttt{MAG\_AUTO} with \texttt{PHOT\_AUTOPARAMS} set to the default values of 2.5 and 3.5. This parameter couplet sets the multiplicative factor and minimum Kron radius used in the ``auto" measurement and is explained in greater detail in the \texttt{Source Extractor} documentation.\footnote{\url{https://sextractor.readthedocs.io/en/latest/Photom.html}} These measurements are reported in Table \ref{tab:host_info_dets}.  We convert these apparent magnitudes to absolute UV magnitudes at 1600~\AA\ using the distance modulus and a $K$-correction, as detailed below.  We first aperture correct the apparent magnitudes using the Encircled Energy (EE) tables from STScI.\footnote{\url{https://www.stsci.edu/hst/instrumentation/wfc3/data-analysis/photometric-calibration/ir-encircled-energy}}   We interpolate the table values for F110W and F140W with a cubic spline to determine the appropriate EE term for the precise \texttt{KRON\_RADIUS} used by \texttt{Source Extractor} for each galaxy.  We then correct these aperture-corrected magnitudes for Galactic dust absorption as reported in the NASA/IPAC Extragalactic Database (NED; \citealt{SF2011}) at the location of the afterglow.  We assume a UV spectral slope of $\beta = -2$ (see Figure 2 in \citealt{Wilkins2013}), where $f_{\nu} \propto \nu^{-\beta}$ and then apply a $K$-correction of $-2.5\log_{10}(1+z)$.  The component of the $K$-correction for the spectral shape is proportional to $(2+\beta)$, and therefore vanishes since we assume $\beta = -2$.  In summary, \begin{equation*}\begin{split}
    M_{1600\mathrm{\AA}} = m_{F110W} + 2.5\log_{10}(EE_{\rm frac}) - A_{\rm MW} + 5 \\ - 5\log_{10}(D_{L}/pc) + 2.5\log_{10}(1+z),
\end{split}
\end{equation*} 
where $D_L$ is the luminosity distance.  We report in Table \ref{tab:host_info_dets} absolute magnitude, $M_{1600\mathrm{\AA}}$, uncertainties as the uncertainty on the apparent magnitude as reported by \texttt{Source Extractor} with propagation of the redshift uncertainty, when reported.

We report $3\sigma$ lower limits on the observed magnitude for sources that are not detected in our images.  In each \textit{HST} image of a non-detected galaxy, we measure the flux within randomly-placed $0.37''$-radius apertures within $6''$ of position of the afterglow. This aperture size was chosen as it is the average radius used for the detections.  We calculate the median flux within these regions, and we clip any flux measurements with a $>3\sigma$ divergence from this value and then recalculate the median. We repeat this $3\sigma$ median-clipping until convergence of the median.  Three standard deviations above this median value is used as an upper limit for the magnitude of the host galaxy.  We then aperture correct these limits using the same methods as were used for the detections and report these final upper limits in Table \ref{tab:host_info_lims}.  These galaxy magnitudes, both detections and upper limits, were derived in this way for modeling and comparison of the UV luminosity function (LF) of our sample, which we detail in Section \ref{sec:discussion}.

\begin{longrotatetable}
\begin{deluxetable*}{lllrrrrlr}
    \tablecaption{List of GRBs in our sample and their afterglow localizations.  From left to right: GRB name, position and uncertainty of the afterglow (as measured from afterglow imaging or reported in the literature), redshift of the afterglow, filter of the afterglow imaging, and references for the afterglow images (or reported position). Afterglow redshift citations are in the Appendix (\ref{sec:appendix}).\\$^*$ Positions for these afterglows are reported from the literature. \label{tab:afterglow_info}}
        \tablehead{\colhead{GRB} & \colhead{RA (ICRS, J2000)} & \colhead{Dec (ICRS, J2000)} & \colhead{Unc('') - $1\sigma$} & \colhead{Redshift} & \colhead{Imaging Source} & \colhead{Filter} & \colhead{Date of Imaging} & \colhead{Reference}}
        \startdata
             050502B & 9:30:10.0703 & +16:59:47.177 & 0.060 & $5.2^{+0.3}_{-0.3}$ & TNG & \it{I} & 2005 Mar 03 & \cite{050502B_TNG}\\
             050505 & 09:27:03.2887 & +30:16:23.907 & 0.050 & 4.275  & Keck/LRIS & \it{I} & 2005 Mar 06 &\cite{050505_Keck_LRIS}\\
             050803$^*$ & 23:22:37.84 & +05:47:08.4 & 1.4 & $4.3^{+0.60}_{-2.40}$ & \textit{Swift}/XRT & - & - & \cite{GoadXRT}\\
             050814 & 17:36:45.3814 & +46:20:21.562 & 0.257 & $5.77^{+0.12}_{-0.12}$ & P60 & \it{i} & 2005 Aug 15 &\cite{050814_P60}\\
             050922B$^*$ &  00:23:13.37 & $-$05:36:17.3 & 2 & $4.9^{+0.3}_{-0.6}$ & \textit{Swift}/XRT & - & - & \cite{GoadXRT}\\
             060206 & 13:31:43.4556 & +35:03:03.186  & 0.067 & 4.059 & P60 & Clear & 2006 Feb 06 & \cite{060206_P60}\\
             060223 & 03:40:49.5661 & $-$17:07:48.357 &  0.077 & 4.406 &\textit{Swift}/UVOT & \it{V}  & 2006 Feb 23 &\cite{060223_Swift}\\
             060510B & 15:56:29.6236 & +78:34:12.102  & 0.094 & 4.942 & Gemini/GMOS-N & \it{i}  & 2006 May 10 &\cite{060510B_GMOS_N_Paper}\\
             060522 & 21:31:44.8367 & +02:53:09.607 & 0.054 & 5.11  & TNG & \it{R} & 2006 May 22 &\cite{060522_TNG}\\
             060927 & 21:58:11.9907 & +05:21:48.355 & 0.128 & 5.467  & VLT/FORS2 & \textit{I} & 2006 Sep 30 &\cite{Ruiz-Velasco2007}\\
             071025 & 23:40:17.0849 & +31:46:42.857 & 0.263 & $4.8^{+0.4}_{-0.4}$ & REM & \it{H} & 2007 Oct 25 &\cite{071025_REM}\\
             090516A & 09:13:02.5973 & $-$11:51:15.055 & 0.023 & 4.111 & VLT/FORS2 & \it{R} & 2009 May 17 & \cite{091516A_VLT}\\
             100219A & 10:16:48.4822 & $-$12:34:00.587 & 0.036 & 4.667  & Gemini/GMOS-S & \it{r} & 2010 Feb 20 &\cite{100219A_GMOS_S}\\
             100513A & 11:18:26.8480 & +03:37:39.899 & 0.022 & 4.772  & Gemini/GMOS-N & \it{i} & 2010 May 13 &\cite{100513A_GMOS_N}\\
             111008A & 04:01:48.2508 & $-$32:42:33.260 & 0.080 & 4.99  & Gemini/GMOS-S & \it{R} & 2011 Oct 09 &\cite{111008A_GMOS_S}\\
             120712A & 11:18:21.2254 & $-$20:02:01.292 & 0.058 & 4.175  & Gemini/GMOS-S & \it{R} & 2012 Jul 12 &\cite{120712A_GMOS_S}\\
             130606A & 16:37:35.1301 & +29:47:46.538 & 0.026 & 5.913  & Gemini/GMOS-N & \it{i} & 2013 Jun 07 & \cite{Chornock2013}\\
             131117A & 22:09:19.3354 & $-$31:45:44.477 & 0.084 & 4.042  & VLT/X-Shooter & \it{R} & 2013 Nov 17 & \cite{131117A_xshooter}\\
             140304A$^*$ & 02:02:34.17 & +33:28:26.01 & 0.02 & 5.283 & VLA & - & - &\cite{140304A_VLA}\\
             140311A &  13:57:13.2771 & +00:38:31.388 & 0.060 & 4.954 & Gemini/GMOS-N & \it{i}  & 2014 Mar 12 & \cite{140311A_GMOS_N}\\
             140428A & 12:57:28.4075 & +28:23:06.280 & 0.066 & $4.68^{+0.52}_{-0.18}$  & Keck/LRIS & \it{I}  & 2014 Apr 29 & \cite{140428A_LRIS}\\
             140518A & 15:09:00.6009 & +42:25:05.886 & 0.046 & 4.7055  & Gemini/GMOS-N & \it{i} & 2014 May 18 &\cite{140518A_GMOS_N}\\
             140614A & 15:24:40.4961 & $-$79:07:43.255 & 0.349 & 4.233  & VLT/X-Shooter & \it{i'} & 2014 Jun 14 & \cite{140614A_XShooter}\\
        \enddata
\end{deluxetable*}
\end{longrotatetable}


\begin{deluxetable*}{lrrrrr}
        \tablecaption{\textbf{Photometry of Host Galaxy Detections.} From left to right: name of the GRB, host centroid position in ICRS, apparent magnitude of the host galaxy as reported from \texttt{Source Extractor}, Galactic extinction from NED \citep{SF2011}, and the absolute UV magnitude of the host galaxy as converted using the methods described in Section \ref{sec:obs}. The uncertainty on the absolute magnitude also accounts for that in redshift.\\ $^*$ Galactic extinction for 130606A is $A_{F140W}$\label{tab:host_info_dets}}
        \tablehead{\colhead{GRB} & \colhead{RA (J2000)} & \colhead{Dec (J2000)} & \colhead{$m_{F110W}$ (mag AB)} & \colhead{$A_{MW}$ ($A_{F110W}$)} & \colhead{$M_{UV}$ (mag AB)}}
        \startdata
             050505 & 09:27:03.2886 & +30:16:23.988 & 25.95(0.10)& 0.031 & $-$20.42(0.10)\\
             050803 & 23:22:37.8142 & +05:47:08.511 &  26.08(0.11) & 0.067 &$-20.33(^{+1.51}_{-0.39})$\\
             050814 & 17:36:45.3861 & +46:20:21.756 & 25.46(0.03)& 0.069 & $-$21.47(0.06)\\
             050922B & 00:23:13.2809 & $-$05:36:17.513 & 25.37(0.08) & 0.032 & $-21.21(^{+0.34}_{-0.19})$\\
             060206 & 13:31:43.4549 & +35:03:03.208 & 27.56(0.22) & 0.022 & $-$18.67(0.22)\\
             060223 & 03:40:49.5884 & $-$17:07:48.258 & 26.63(0.07) & 0.101 &$-$19.96(0.07)\\
             060510B & 15:56:29.6623 & +78:34:12.065 & 26.05(0.06) & 0.020 & $-$20.58(0.06)\\
             071025 & 23:40:17.0939 & +31:46:42.862 & 26.06(0.10) & 0.065 & $-$20.51(0.25)\\
             090516A & 09:13:02.6094 & $-$11:51:15.152 & 25.04(0.07) & 0.044 & $-$21.24(0.07)\\
             100513A & 11:18:26.8473 & +03:37:39.837 & 26.65(0.15) & 0.048 & $-$19.89(0.15)\\
             111008A & 04:01:48.2556 & $-$32:42:33.164 & 27.69(0.30) & 0.005 & $-$18.87(0.30)\\
             120712A & 11:18:21.2274 & $-$20:02:01.369 & 27.06(0.12) & 0.037 & $-$19.26(0.12)\\
             130606A & 16:37:35.1338 & +29:47:46.549 & 26.79(0.05) & 0.015$^*$ & $-$20.26(0.05)\\
             140311A & 13:57:13.2765 & +00:38:31.414 & 28.38(0.35) & 0.033 & $-$18.18(0.35)\\
             140518A & 15:09:00.5975 & +42:25:05.708 & 27.22(0.13) & 0.040 & $-$19.31(0.13)\\
             140614A & 15:24:40.5339 & $-$79:07:43.346 & 26.14(0.09) & 0.109 & $-$20.28(0.09)\\
        \enddata
\end{deluxetable*}

\begin{deluxetable*}{lrrr}
        \tablecaption{\textbf{Photometry for Host Galaxy Non-Detections.}  From left to right, apparent magnitudes (as $3\sigma$ above sky and encircled energy corrected), Galactic extinction from NED \citep{SF2011}, and extinction-corrected absolute magnitudes as converted using methods described in Section \ref{sec:obs}.  When applicable, redshift uncertainty was propagated, and the brighter limit was chosen.  \label{tab:host_info_lims}}
        \tablehead{\colhead{GRB}  & \colhead{$m_{F110W}$ (mag AB)} & \colhead{$A_{MW}$ ($A_{F110W})$} & \colhead{$M_{UV}$ (mag AB)}}
    \startdata
        050502B & $> 27.55$ & 0.026 & $>-19.10$\\
        060522 & $> 27.83$ & 0.048 & $>-18.67$\\
        060927 & $> 27.84$ & 0.054 & $>-18.76$\\
        100219A & $> 27.58$ & 0.067 & $>-18.78$\\
        131117A & $> 27.39$ & 0.016 & $>-18.67$\\
        140304A & $> 27.49$ & 0.049 & $>-19.05$\\
        140428A & $> 27.66$ & 0.019 & $>-18.95$\\
    \enddata
\end{deluxetable*}

\begin{figure*}
    \centering
    \includegraphics[width=.19\textwidth]{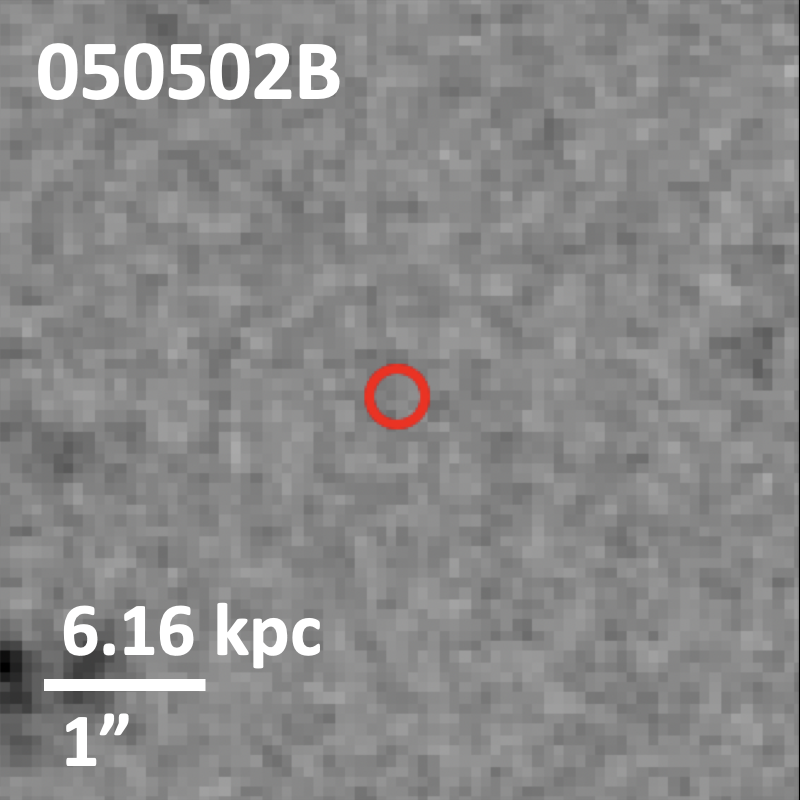}
    \includegraphics[width=.19\textwidth]{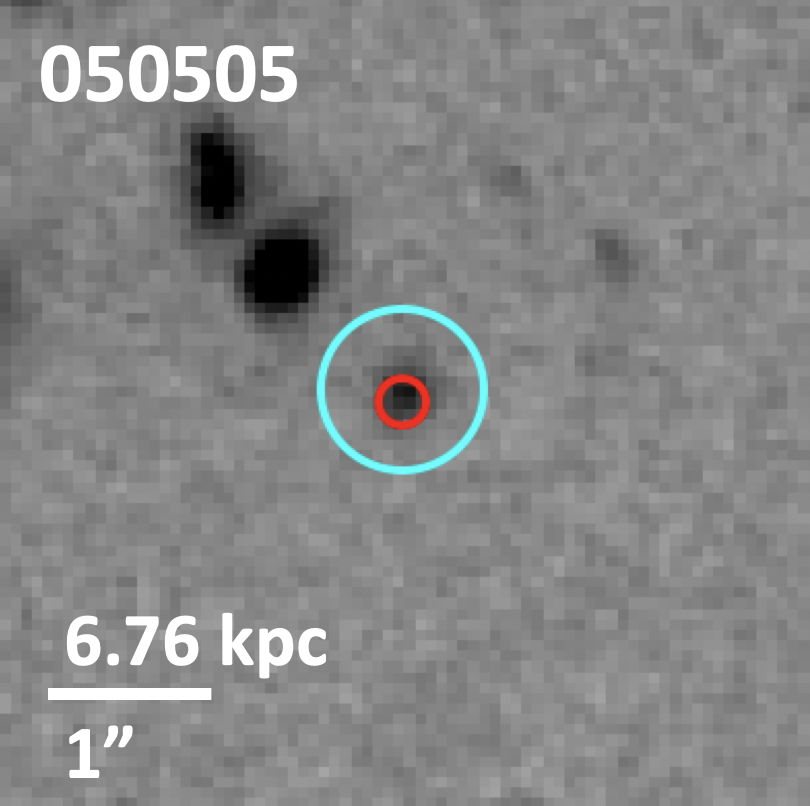} 
    \includegraphics[width=.19\textwidth]{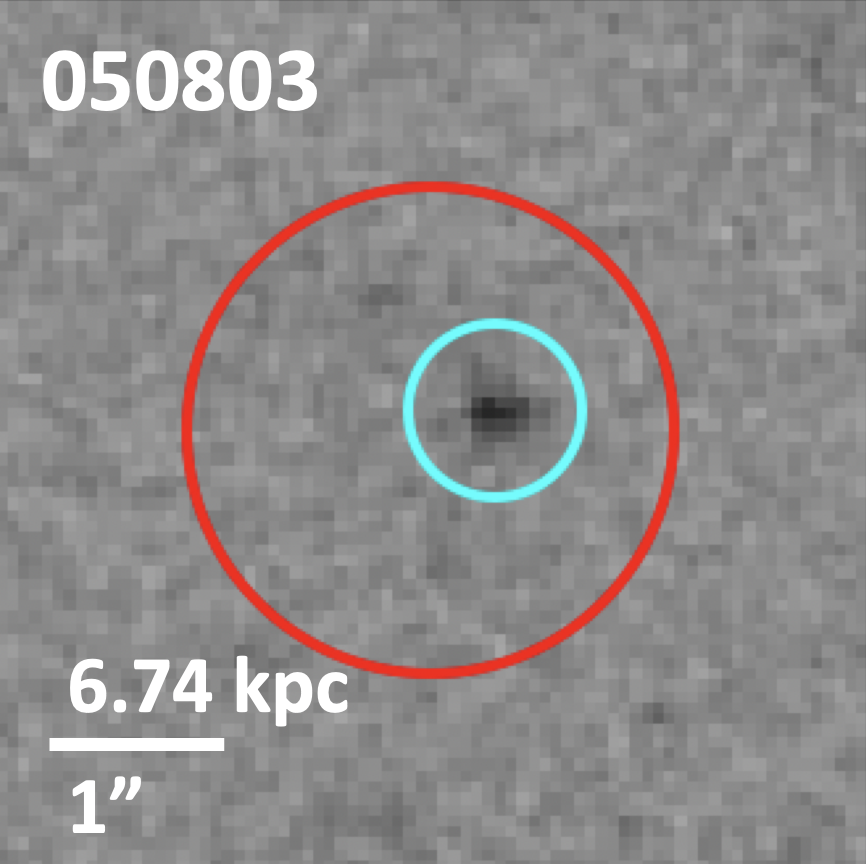}
    \includegraphics[width=.19\textwidth]{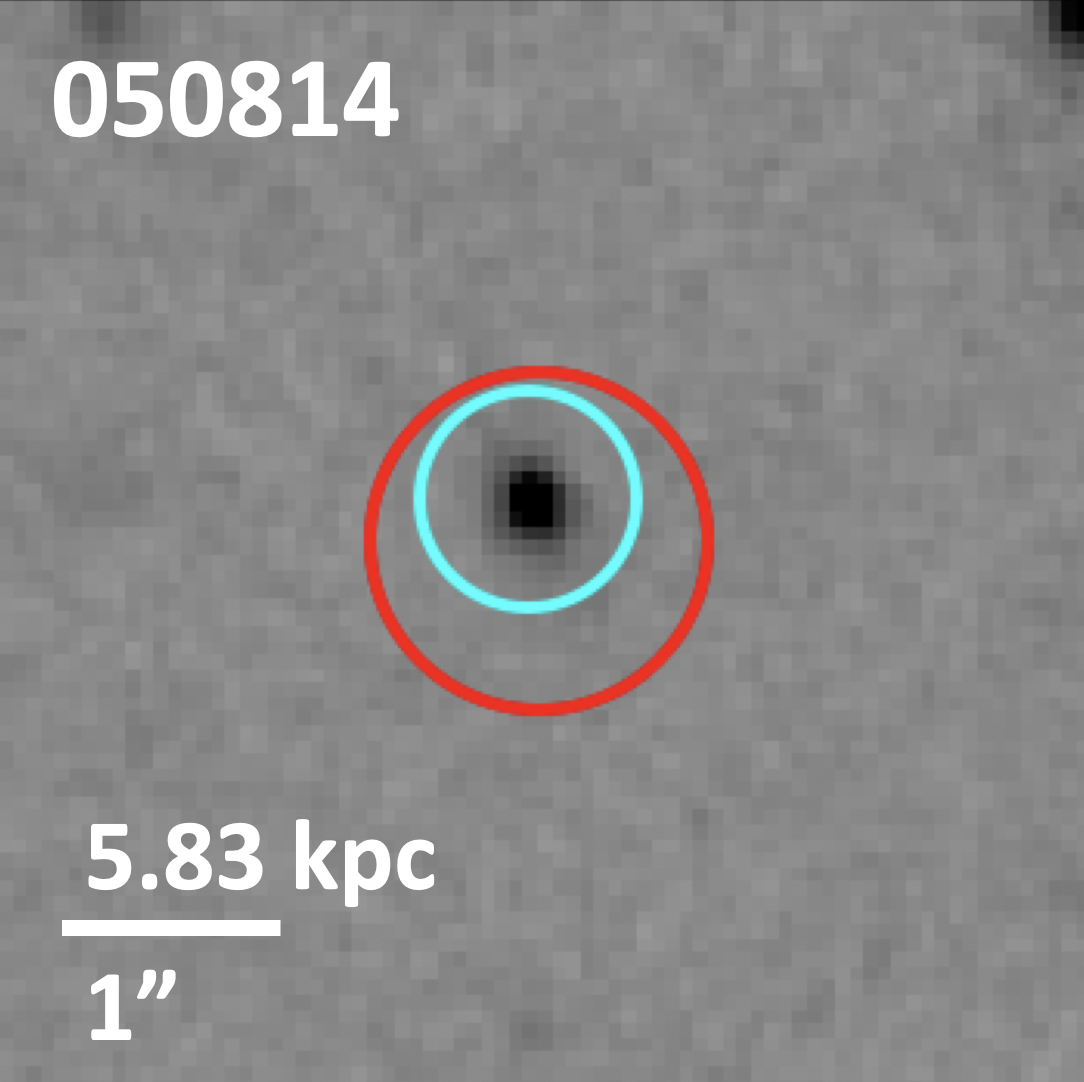} 
    \includegraphics[width=.19\textwidth]{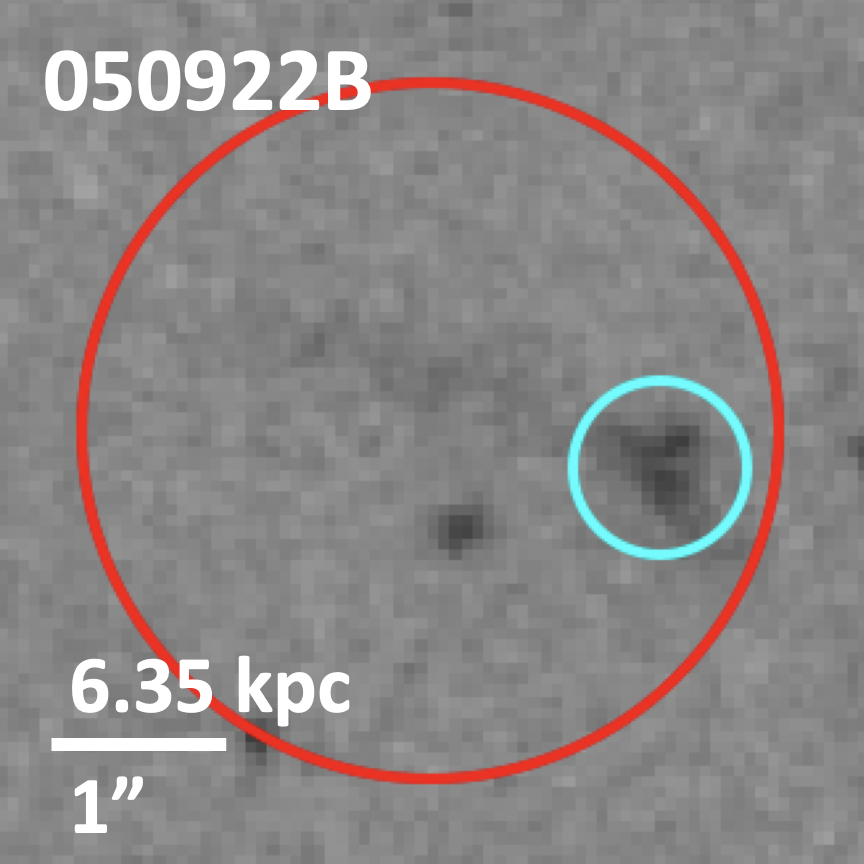}
    \includegraphics[width=.19\textwidth]{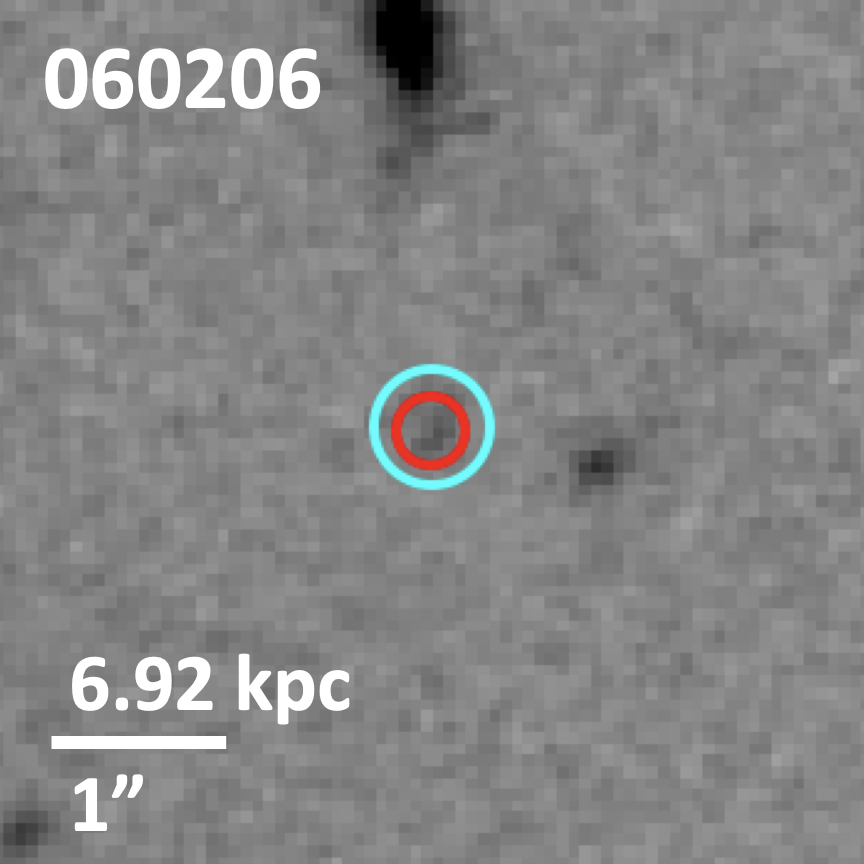}  
    \includegraphics[width=.19\textwidth]{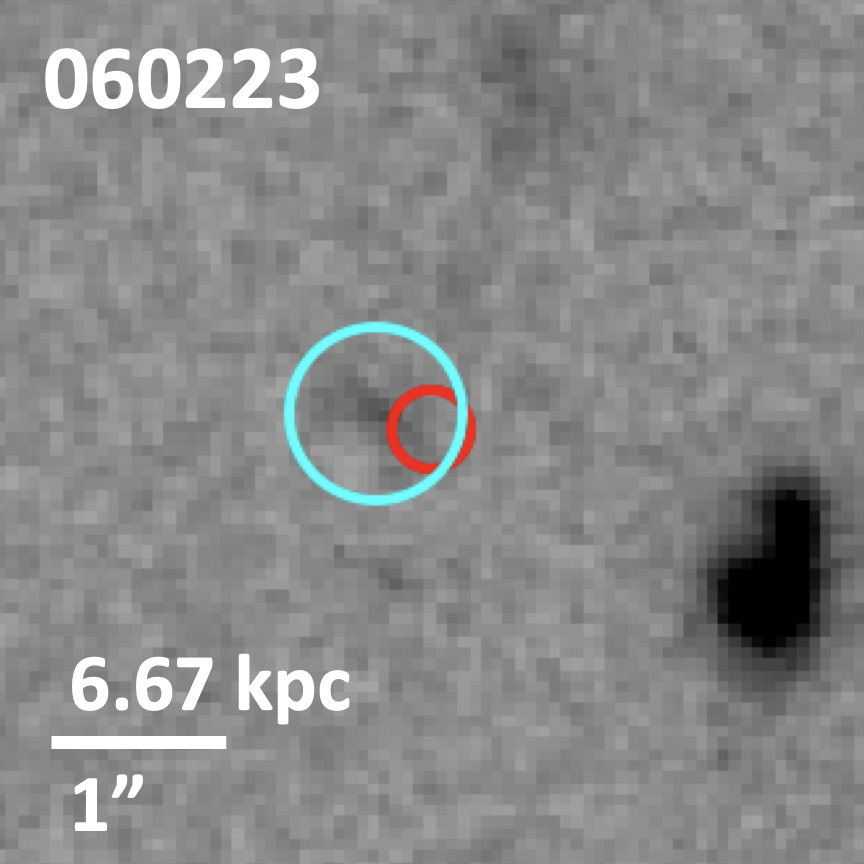}
    \includegraphics[width=.19\textwidth]{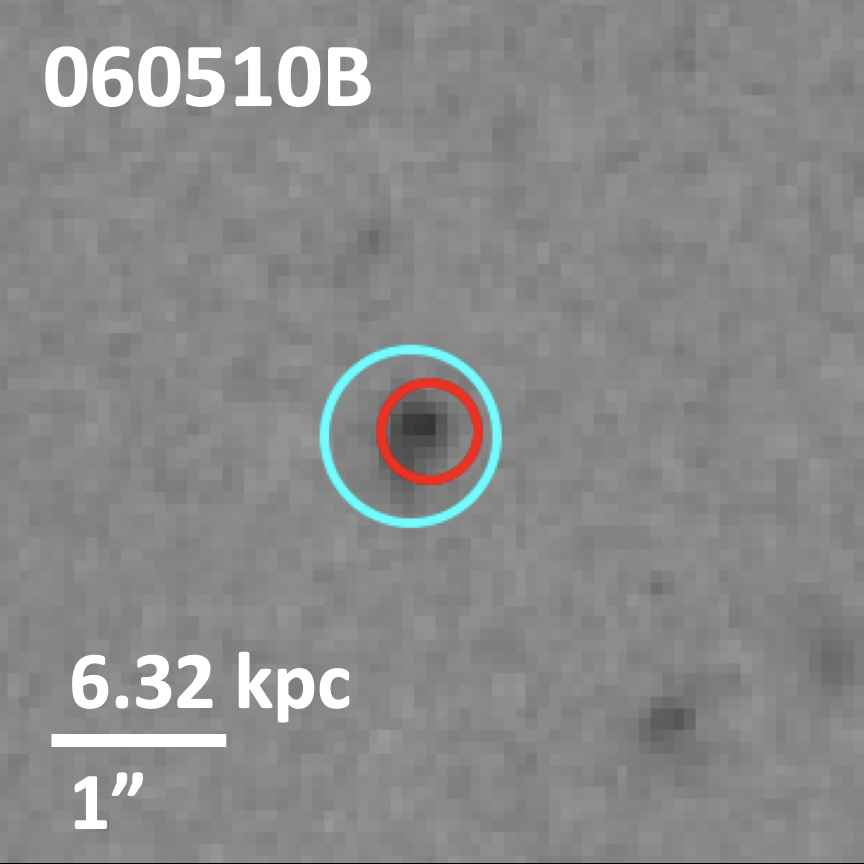}
    \includegraphics[width=.19\textwidth]{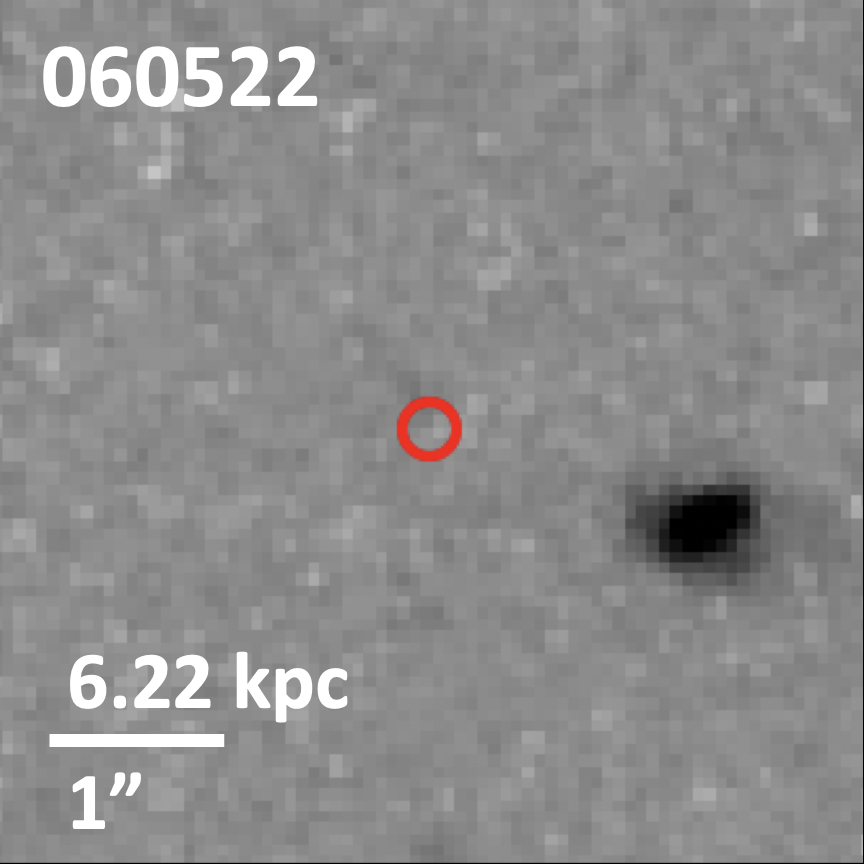}
    \includegraphics[width=.19\textwidth]{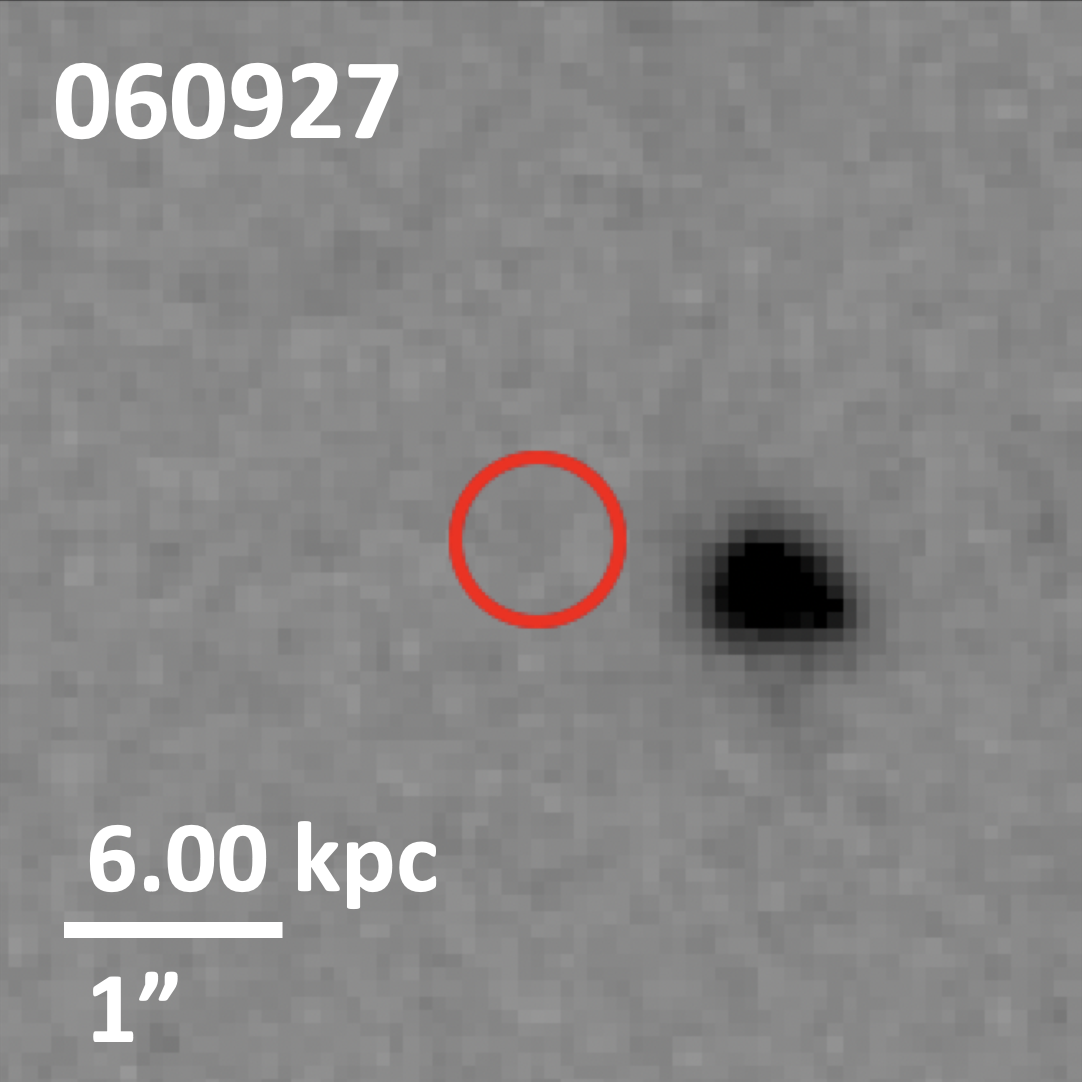} 
    \includegraphics[width=.19\textwidth]{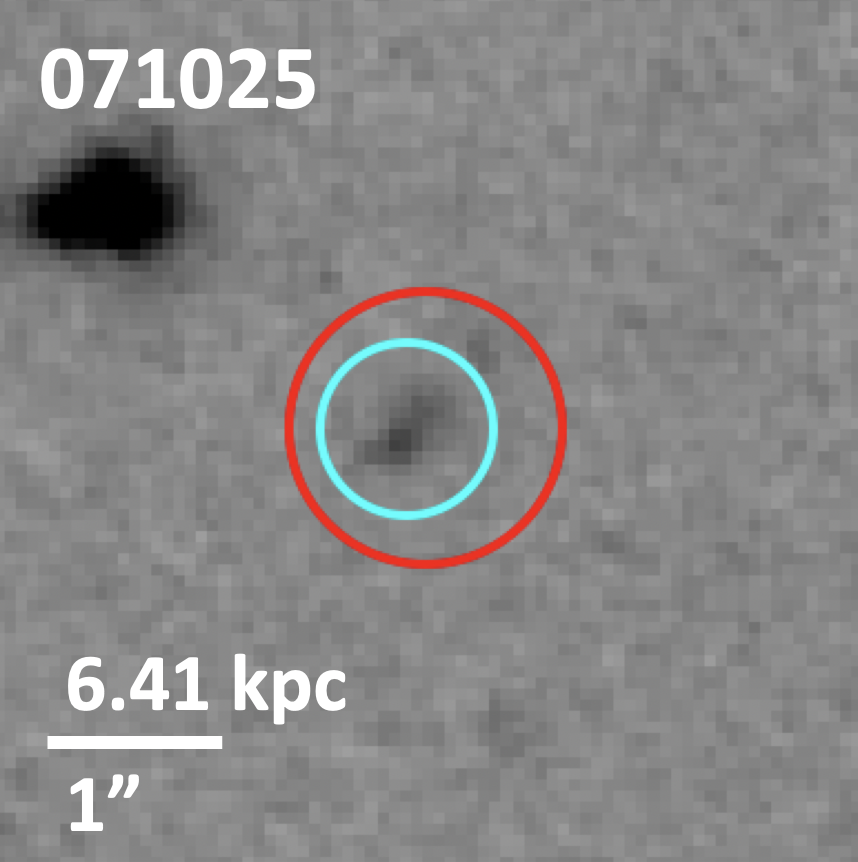}
    \includegraphics[width=.19\textwidth]{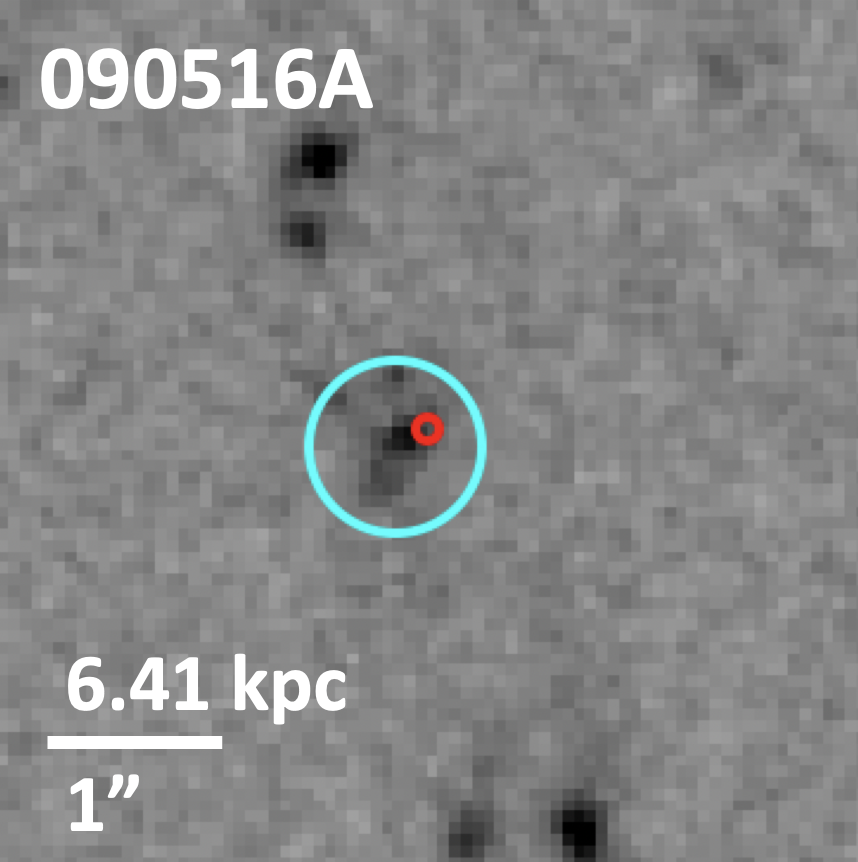} 
    \includegraphics[width=.19\textwidth]{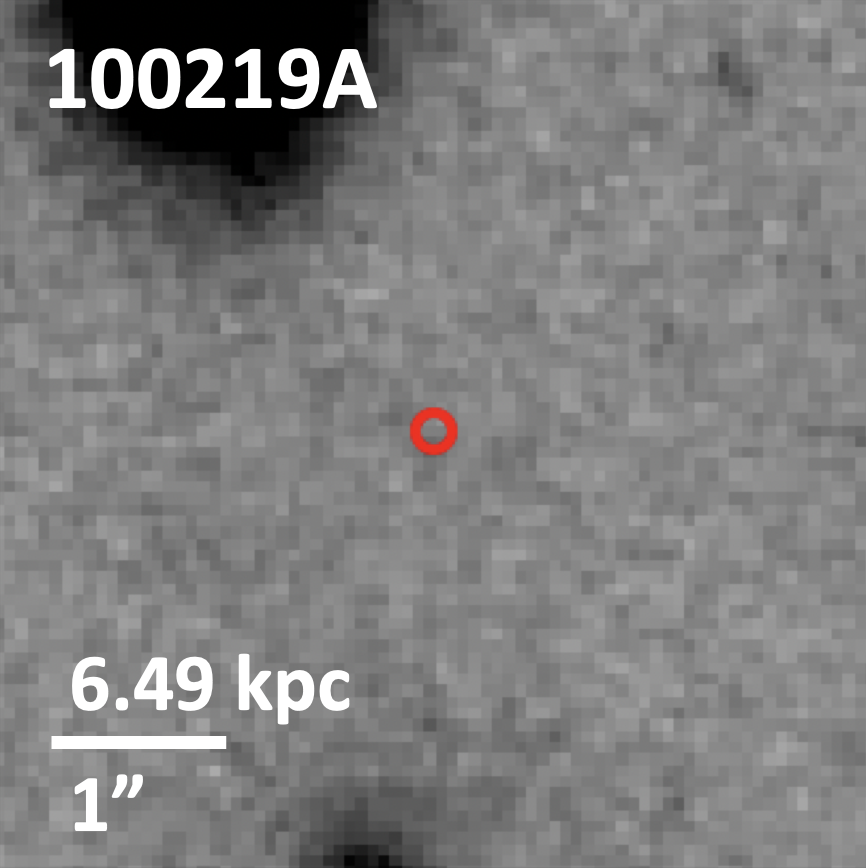} 
    \includegraphics[width=.19\textwidth]{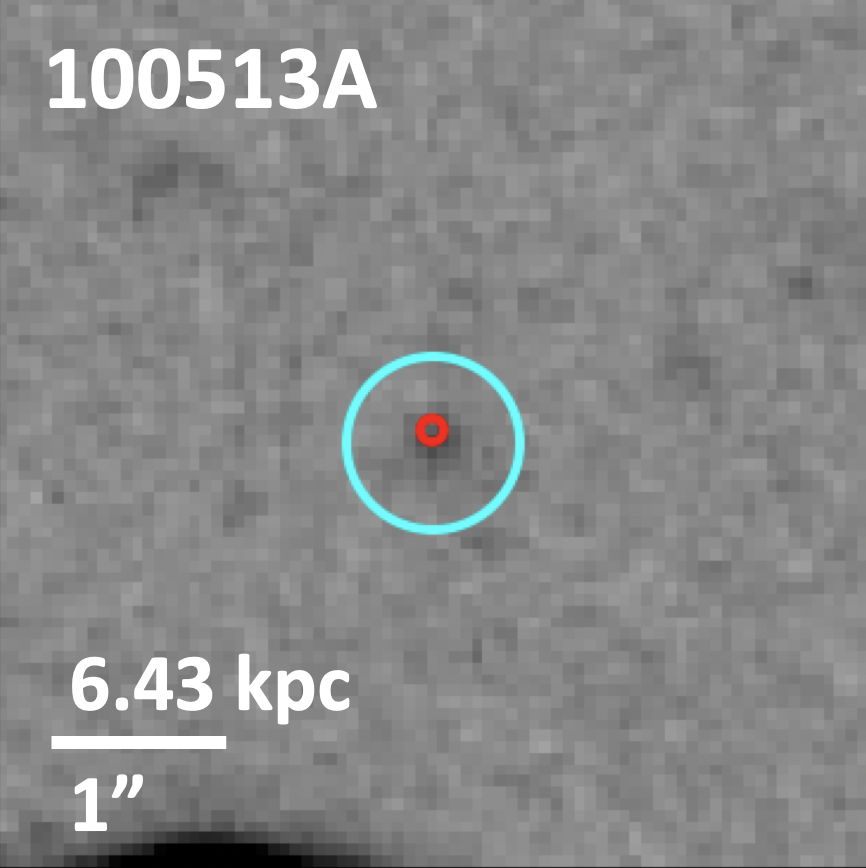}
    \includegraphics[width=.19\textwidth]{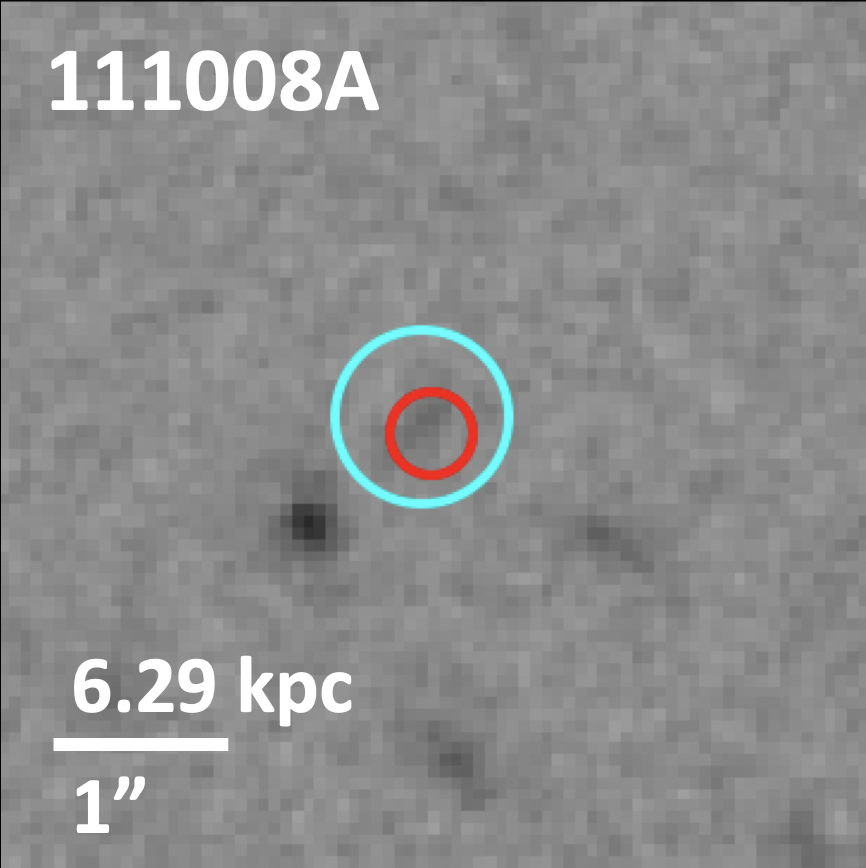} 
    \includegraphics[width=.19\textwidth]{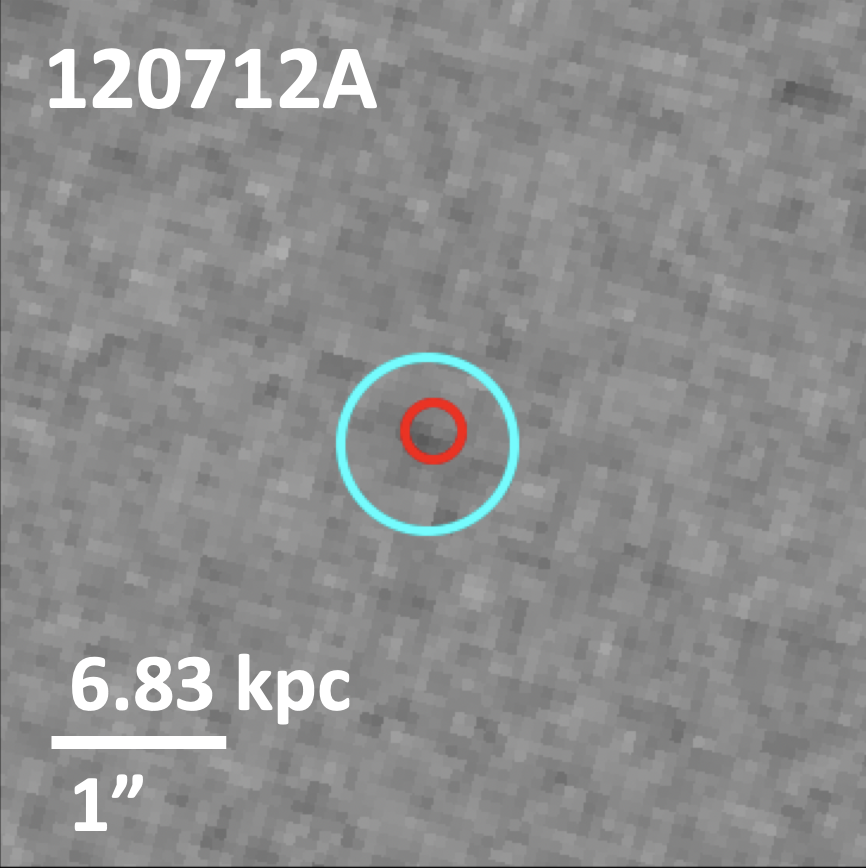}
    \includegraphics[width=.19\textwidth]{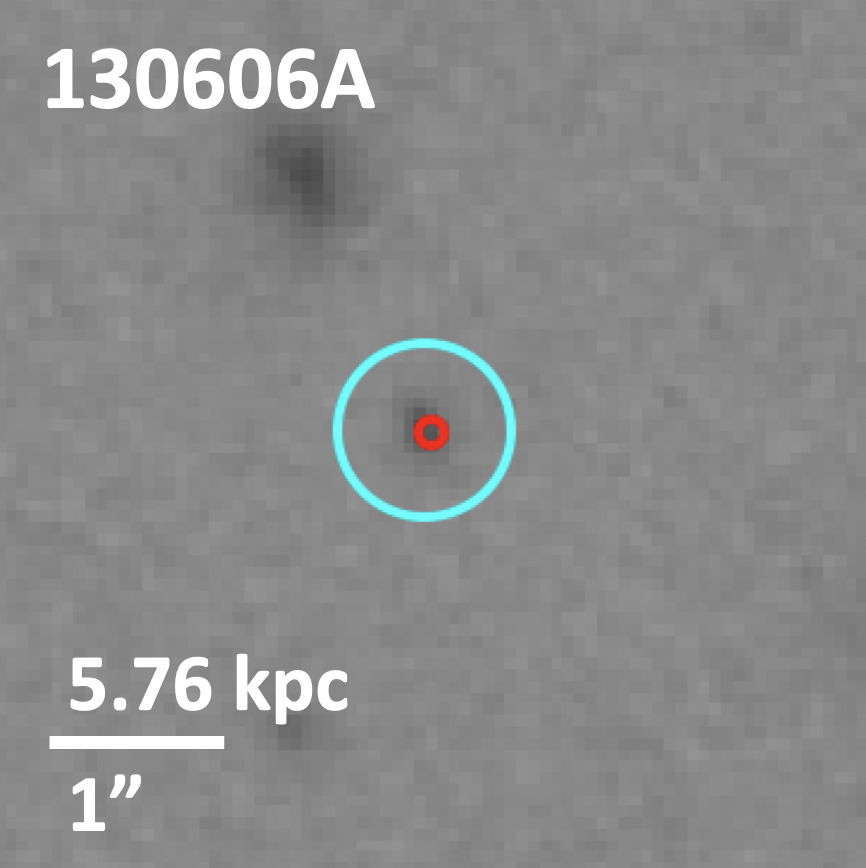}
    \includegraphics[width=.19\textwidth]{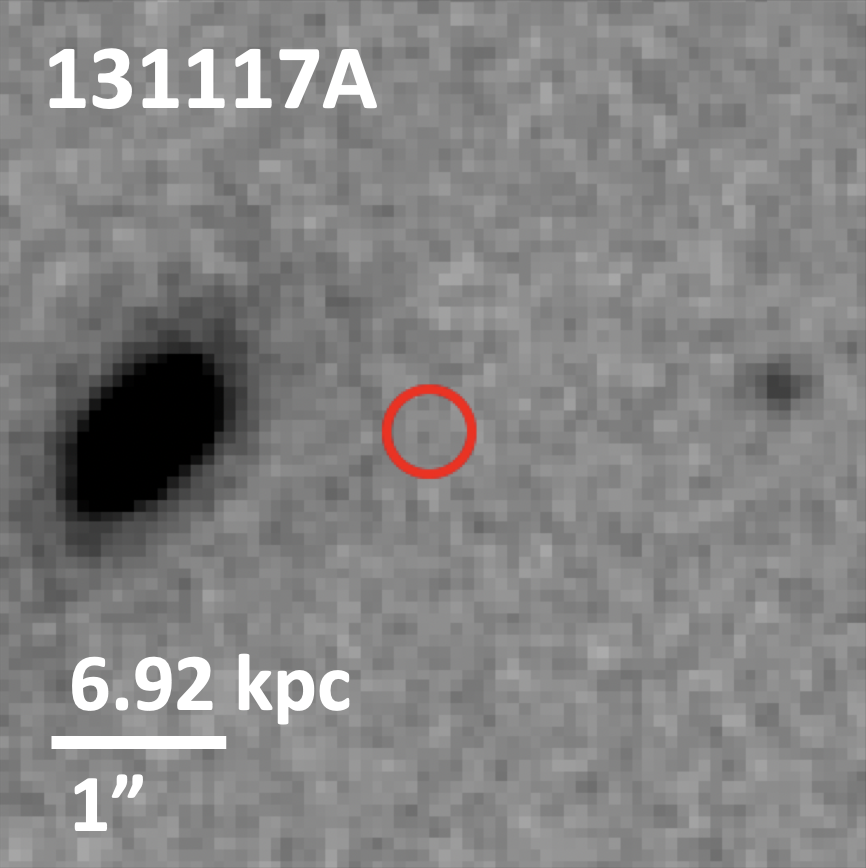}
    \includegraphics[width=.19\textwidth]{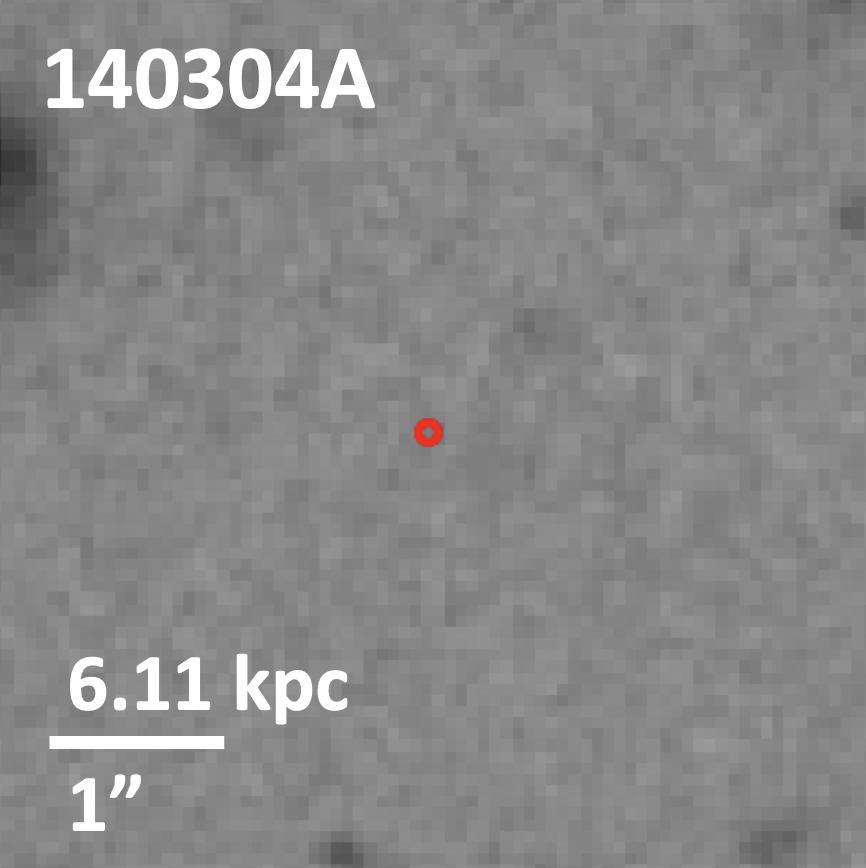}
    \includegraphics[width=.19\textwidth]{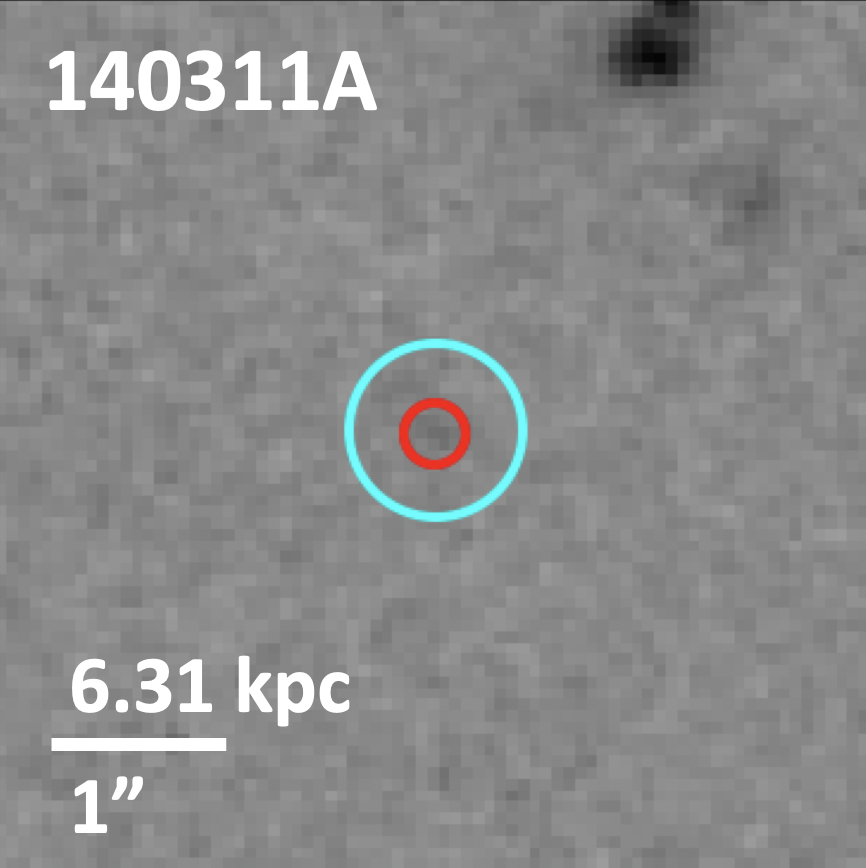}
    \includegraphics[width=.19\textwidth]{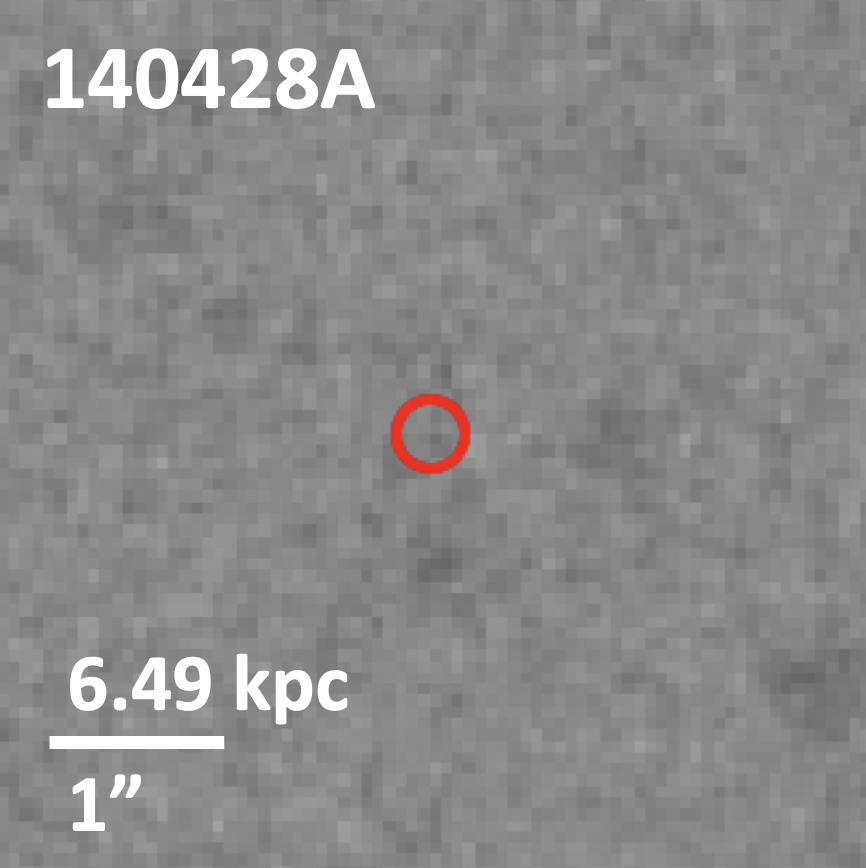} 
    \includegraphics[width=.19\textwidth]{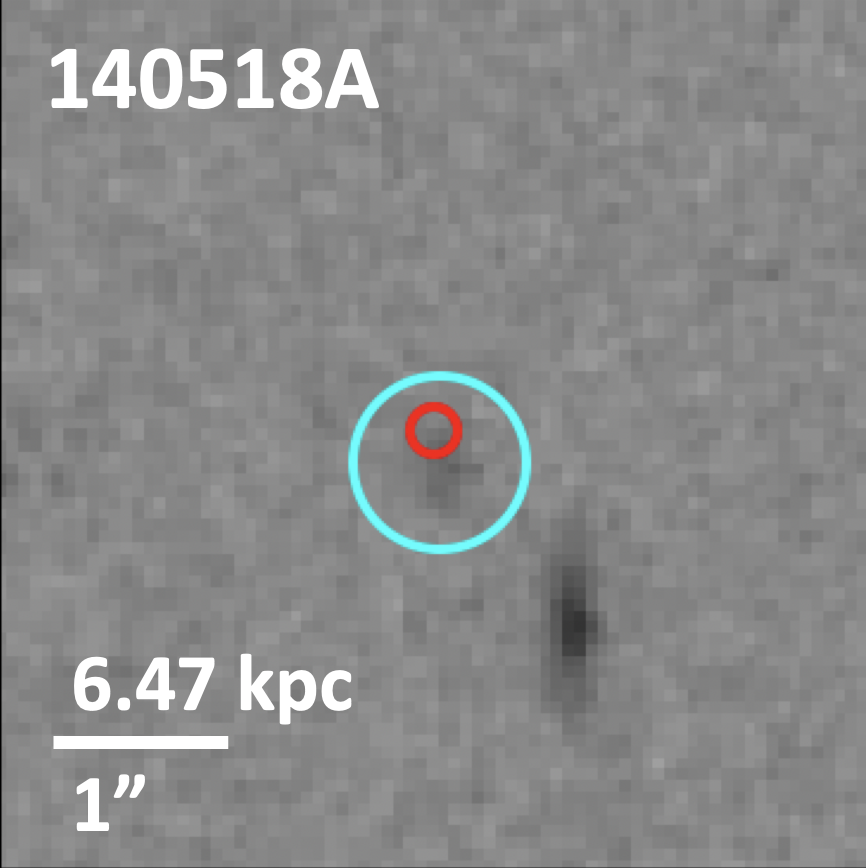}
    \includegraphics[width=.19\textwidth]{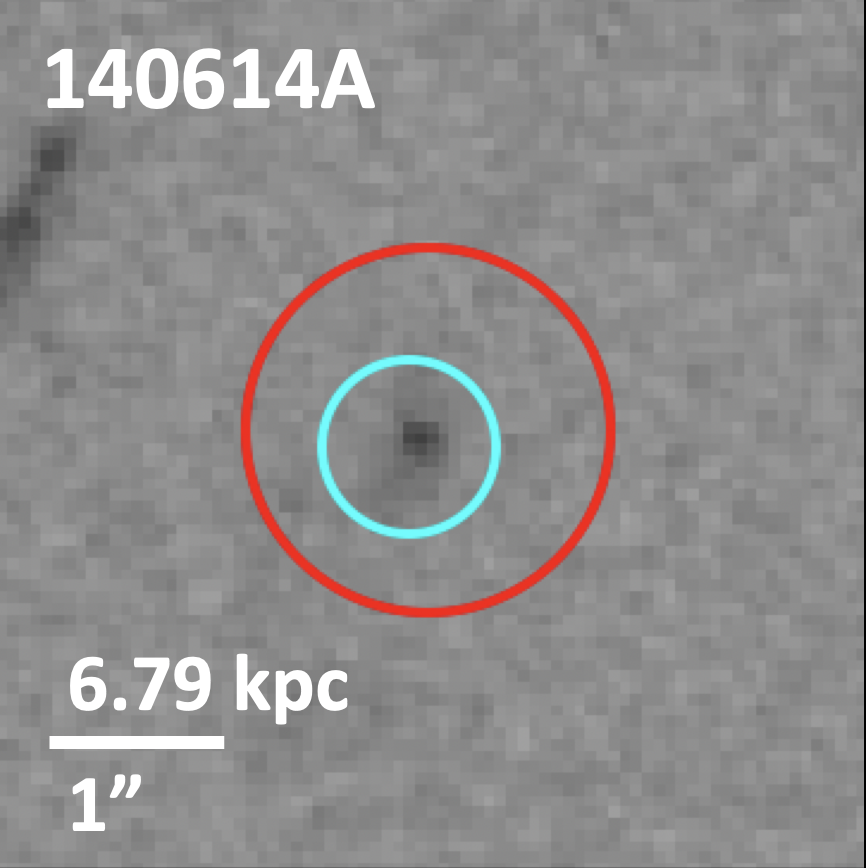}\\
    \caption{IR imaging from WFC3 of the 23 host galaxies.  Each box is $5''$ wide.  The afterglow position is shown in red with either the $3\sigma$-radius or with the \textit{Swift}-XRT radius, while the host (when detected) is identified by a cyan $0.5''$-radius region. Positions of both the afterglow and host are reported in Tables \ref{tab:afterglow_info} and \ref{tab:host_info_dets}, respectively.  North is up and East is to the left.}
    \label{fig:Stamp_Collage}
    \end{figure*}

\section{Discussion}\label{sec:discussion}

\subsection{Lyman-break Galaxy UV Luminosity Functions}
We derive the luminosity distribution of our GRB host galaxy sample from Tables \ref{tab:host_info_dets} and \ref{tab:host_info_lims}.  We compare these results to samples of Lyman-break galaxies at $z\sim5$ from \cite{Bouwens_2021a} and \cite{Finkelstein2015}. We elected to not use results from the more recent \cite{Bouwens2022} due to their choice of functional form for the luminosity function, which deviates from the standard Schechter function by including an additional parameter, $\delta$, that allows for curvature at the faint end ($M_{UV} > -16$ mag) of the luminosity function. Their formula and best fit parameters result in a divergent luminosity function whose CDF is inherently highly sensitive to the choice of the  lower integration limit. Furthermore, since our faintest detected GRB host galaxy is at $M_{UV}=-18.1$ mag and even the upper limits for our non-detections are not much fainter than this, our data are insensitive to the shape of the faint-end LF and could not place any meaningful limitations on this additional parameter.  Indeed, even with their larger sample of 59 $z\sim5$ Lyman-break galaxies, they find $\delta = 0.07 \pm 0.2$, to an uncertainty $5\times$ greater than that they find for their $\alpha\ (0.04)$. The Schechter function parameters from both \cite{Bouwens_2021a} and \cite{Finkelstein2015} are reported in Table \ref{tab:RStan_fit_params}.

To meaningfully compare the Lyman-break galaxy luminosity functions of \citealt{Bouwens_2021a} and \cite{Finkelstein2015} to that of our GRB host galaxies (detailed in following sections), we must first account for the GRB production rate.  To do so, it is necessary to weight the Lyman-break galaxy luminosity functions by the instantaneous star-formation rate (SFR), as the GRB production rate is expected to be proportional to the SFR. The SFR is proportional to the intrinsic UV luminosity \citep{Kennicutt1998}, and so we can effectively account for GRB selection effects by multiplying the Lyman-break galaxy luminosity function by the intrinsic luminosity of the Lyman-break galaxy.  We consider two conversions of the intrinsic to the observed UV luminosity, as the luminosity functions are functions of the observed luminosity. In both cases, we construct the base SFR-weighted Schechter LF (i.e., a predicted GRB host luminosity function) as below:

\begin{equation}\label{eq:1}
    \phi(L_{obs}) = \Bigg(\frac{\phi_*}{L_*}\Bigg)\Bigg(\frac{L_{obs}}{L_*}\Bigg)^{\alpha} e^{-L_{obs}/L_*} \times L_{int}
\end{equation}

\noindent where $L_*$ is the characteristic luminosity, $L_{int}$ is the intrinsic luminosity, $L_{obs}$ is the observed luminosity, $\phi_*$ is a normalization parameter, and $\alpha$ is the faint-end slope, as is standard in the Schechter function.  In magnitude space, this can be restated as:

\begin{equation}\label{eq:2}
    \phi(M_{obs}) = \phi_{**}\Big(10^{f(M_{obs})}\Big)^{(\alpha+1)} e^{-10^{f(M_{obs})}} \times 10^{-0.4\times M_{int}}
\end{equation}

\noindent where $f(M_{obs}) = 0.4\times(M_*-M_{obs})$, $M_{int}$ is the intrinsic magnitude, $M_{obs}$ is the observed magnitude, $\phi_{**}$ is a normalization parameter, and $\alpha$ is still the faint-end slope. $M_*$ is the characteristic magnitude and is defined as $M_* = -2.5\times\log{\frac{L_*}{L_0}}$, where $L_0$ is the luminosity of a source with an absolute magnitude of 0.

\begin{enumerate}
    \item In our first formalism, we assume a luminosity-independent dust-contribution where the intrinsic luminosity, $L_{int}$, is linearly proportional to the observed luminosity, $L_{obs}$. Here, $L_{int} \propto L_{obs} \propto 10^{-0.4\times M_{obs}}$.

    \item Our second formalism is one where we assume a non-linear luminosity-dependent dust-contribution.  We make this assumption because we expect more massive galaxies to have more dust.  We construct  this formalism from the following two relations from \cite{Overzier2011} \footnote{The amount of host UV extinction due to dust at $z\sim5$ is an active area of research.  The choice for this correction has often been that from \cite{Meurer1999}, $A_{1600} = 4.43 + 1.99\beta$.  However, there have been several updates to this relation, e.g., \citep{Overzier2011, Takeuchi2012, Casey2014, Bouwens2014b}.  Here, we elect to use the relation from \cite{Overzier2011} (as stated in Eq. \ref{eq:3}), as it is measured from Lyman-break analog galaxies, which offers the closest comparison to our GRB hosts.} and \cite{Bouwens2014b}, respectively:

\begin{equation}\label{eq:3}
    A_{1600} = 1.81\beta+4.01,
\end{equation}

where $A_{1600}$ is the extinction at 1600~\AA\ and $\beta$ is defined as: 

\begin{equation}\label{eq:4}
    \beta = -1.91 - 0.14(M_{UV} + 19.5).
\end{equation}

Since $A_{1600}$ cannot be negative, this results in a piecewise luminosity function of the form of Eq. (\ref{eq:1}) where now 

\begin{equation}\label{eq:5}
L_{int} \propto 10^{-0.4\times M_{obs}} \text{ for }M_{obs} > -17.3
\end{equation}

and 
\begin{equation}\label{eq:6}
L_{int} \propto 10^{-0.4\times(1.25\times M_{obs}+4.39)} \text{ for } M_{obs}\leq 17.3
\end{equation}

\item Our third formalism is one where we again assume a non-linear luminosity-dependent dust contribution, however we substitute for Eq. (\ref{eq:3}) an estimation of the same relation from \cite{Meurer1999}:

\begin{equation}\label{eq:7}
    A_{1600} = 1.99\beta+4.43
\end{equation}

\end{enumerate}

We refer to our second and third formalisms as ``O11" and ``M99," respectively, in reference to the choice of the $A_{1600}(\beta)$ formulation (i.e., the choice of Eq. (\ref{eq:3}) or Eq. (\ref{eq:7})).

\subsection{GRB Host UV Luminosity Function}

We use Bayesian hierarchical modelling to constrain the parameters of the luminosity distribution of the GRB host galaxies. We assume the galaxies follow a SFR-weighted Schechter function (see Eq. \ref{eq:1}) with a faint-end magnitude limit of $M_{UV} = -14.1$ mag (this arbitrary magnitude choice converts to a convenient value in our luminosity units and is well below our detection threshold in all cases, although we find that our results are not statistically sensitive to this precise choice). We used weakly informative Gaussian priors of $\mu_{\alpha} = -1.6, \sigma_{\alpha} = 1.0$ and $\mu_{\log L_{*}} = 10, \sigma_{\log L_*} = 1 $ for the $\alpha$ and $ log L_{*}$ parameters, respectively. The model self-consistently included both the detections and the seven upper limits: the luminosity for each of these 23 objects was a free parameter in the model, and hence each has a corresponding posterior distribution. We symmetrized the uncertainties for each measurement, conservatively selecting the greater of the two, though we find that our results are also not sensitive to this choice. Four chains were run per model with at least 100,000 samples per chain after warm-up, which ensured negligible MCMC standard errors for all parameters of interest. In the final model runs, there were no divergences, and the chains for all parameters mixed well, with the convergence diagnostic $\widehat{R}=1$.  We complete this process three times, once each for our different considerations of the SFR-weight on the Lyman-break galaxy LF as described in the previous section. To model these luminosity distributions of the GRB host galaxies, we use the \texttt{Stan} software as implemented in version 2.26.13 of the {\tt RStan} package \citep{RStan_citation}.  

\begin{figure*}
    \centering
    \includegraphics[width=0.49\textwidth]{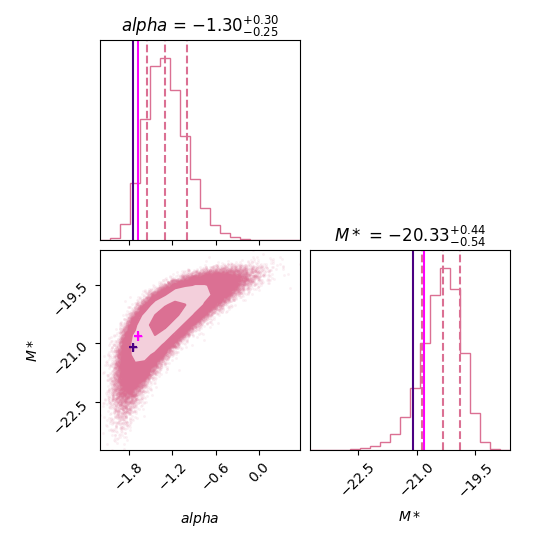}
    \includegraphics[width=0.49\textwidth]{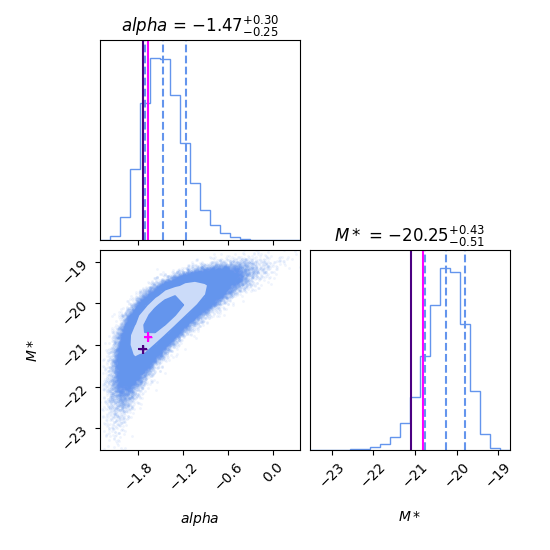} 
    \includegraphics[width=0.49\textwidth]{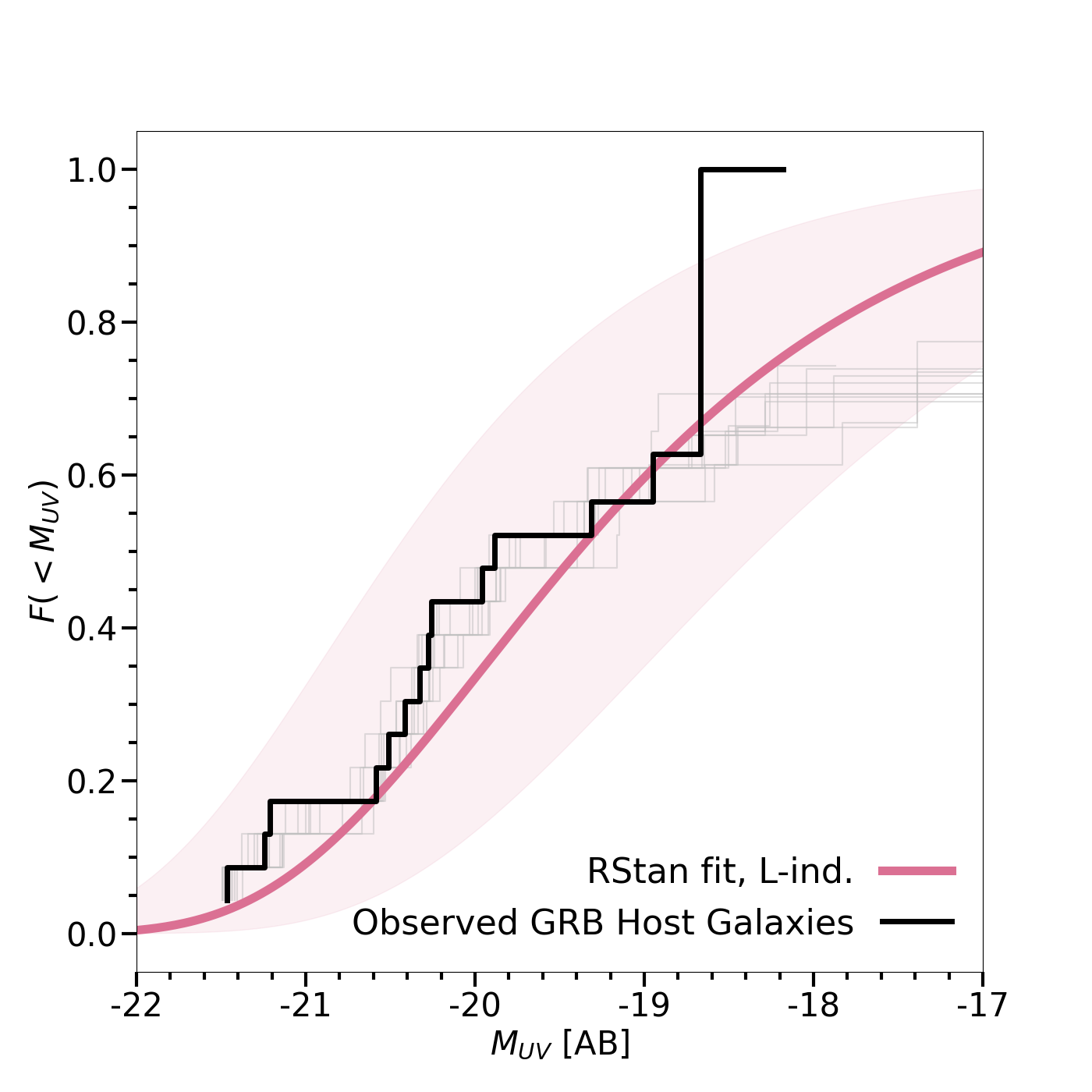}
    \includegraphics[width=0.49\textwidth]{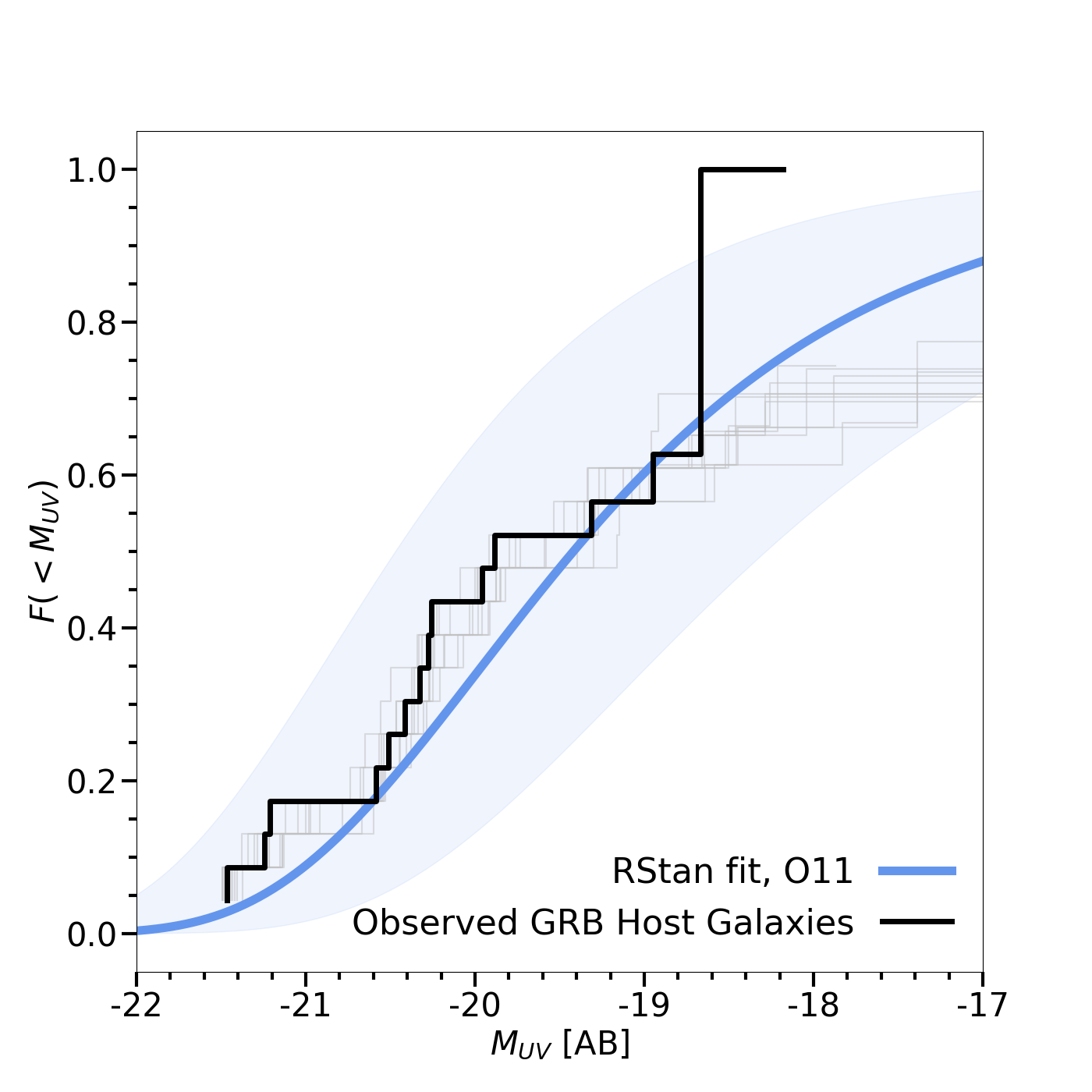} 
     
    \caption{\textbf{Top:} Correlations and marginalized posterior densities for $\alpha$ and $M_*$ in the SFR-weighted UV luminosity function for the $L$-independent (pink) and O11 (blue) extinction corrections.  The indigo lines show the $\alpha$ and $M_*$ parameters from \cite{Bouwens_2021a}, while the fuchsia lines show the same parameters for \cite{Finkelstein2015}.  These parameters are also shown with their uncertainty as crosses in the center panels.  $1\sigma$ and $2\sigma$ contours are shown in the 2D histograms.  \textbf{Bottom:} The observed cumulative distribution functions (CDFs) of and best fits to the UV luminosity function for GRB host galaxies at $z\sim5$. As indicated in the legend, the model shown in pink assumes a $L$-independent extinction correction and the model in blue assumes the O11 extinction correction.  The black line is the observed GRB host galaxy CDF. The same 10 random draws from the modeled data are shown in silver.}
    \label{fig:rstanLFs}
\end{figure*}

We show the posteriors and best-fit SFR-weighted Schechter functions for the $L$-independent and O11 weightings in Figure \ref{fig:rstanLFs}.  These best-fit $\alpha$ and $M_*$ parameters, along with their $1\sigma$ uncertainties, are provided in Table \ref{tab:RStan_fit_params} as well as the same parameters for the M99 weighting.  The Schechter parameters from  \cite{Bouwens_2021a} ($\alpha = -1.74 \pm 0.06$, $M_* = -21.10 \pm 0.11$ mag) are consistent to within $2\sigma$ to our O11 best fit parameters ($\alpha = -1.47 \pm 0.27$, $M_* = -20.25 \pm 0.47$ mag).  The same is true for the parameters from \cite{Finkelstein2015}.  These Lyman-break galaxy fits are consistent to within $2\sigma$ to our $L$-independent weighting as well.  The slightly better agreement of the Lyman-break galaxy LFs to the O11 formalism is expected, as this formalism offers a more realistic estimate of the intrinsic extinction at $z\sim5.$  Along this parametric comparison, there is no evidence of disagreement between the GRB host galaxy sample and the Lyman-break galaxy samples.

While the differences between the galaxy samples are not statistically significant, the best fits to the GRB host galaxies have a shallower $\alpha$ and a fainter $M_*$.  With a larger GRB host galaxy sample, if these parameter differences were to become statistically significant, the move toward a shallower $\alpha$ and a fainter $M_*$ would indicate that Lyman-break galaxy LFs over-predict the amount of faint star-formation.

\begin{deluxetable}{lrr}
        \tablecaption{Best fit Schechter function fit parameters to the GRB host data and the Lyman-break galaxy data sets. The fits to the GRB hosts were measured from our \texttt{RStan} program, while fits to the Lyman-break galaxies were copied from \cite{Bouwens_2021a} and \cite{Finkelstein2015}.
        \label{tab:RStan_fit_params}}
        \tablehead{\colhead{Formalism} &  \colhead{$\alpha$} & \colhead{$M_*$ (mag)}}
        \startdata
            Linear L-conversion &  $-1.30^{+0.30}_{-0.25}$ & $-20.33^{+0.44}_{-0.54}$ \\
            O11 formalism & $-1.47^{+0.30}_{-0.25}$ & $-20.25^{+0.43}_{-0.51}$\\
            M99 formalism & $-1.49^{+0.30}_{-0.25}$&$-20.25^{+0.42}_{-0.51}$ \\
            \hline
            Lyman-break galaxy Samples \\
            \hline
            B21a & $-1.74^{+0.06}_{-0.06}$ & $-21.10_{-0.11}^{+0.11}$\\
            F15 & $-1.67_{-0.06}^{+0.05}$ & $-20.81_{-0.12}^{+0.12}$\\
        \enddata
\end{deluxetable}

\begin{figure*}
    \centering
    \includegraphics[width=0.475\textwidth]{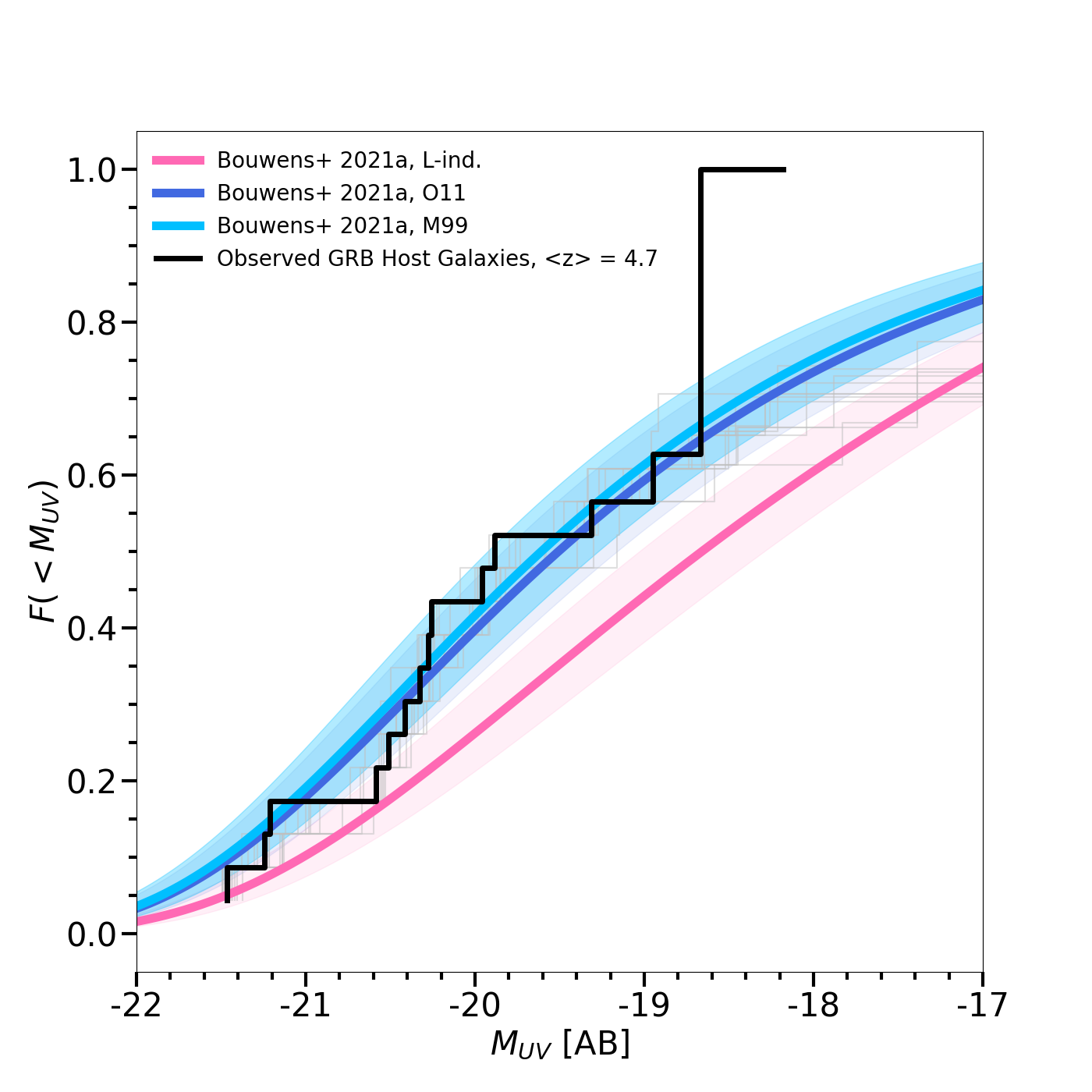} 
    \includegraphics[width=0.475\textwidth]{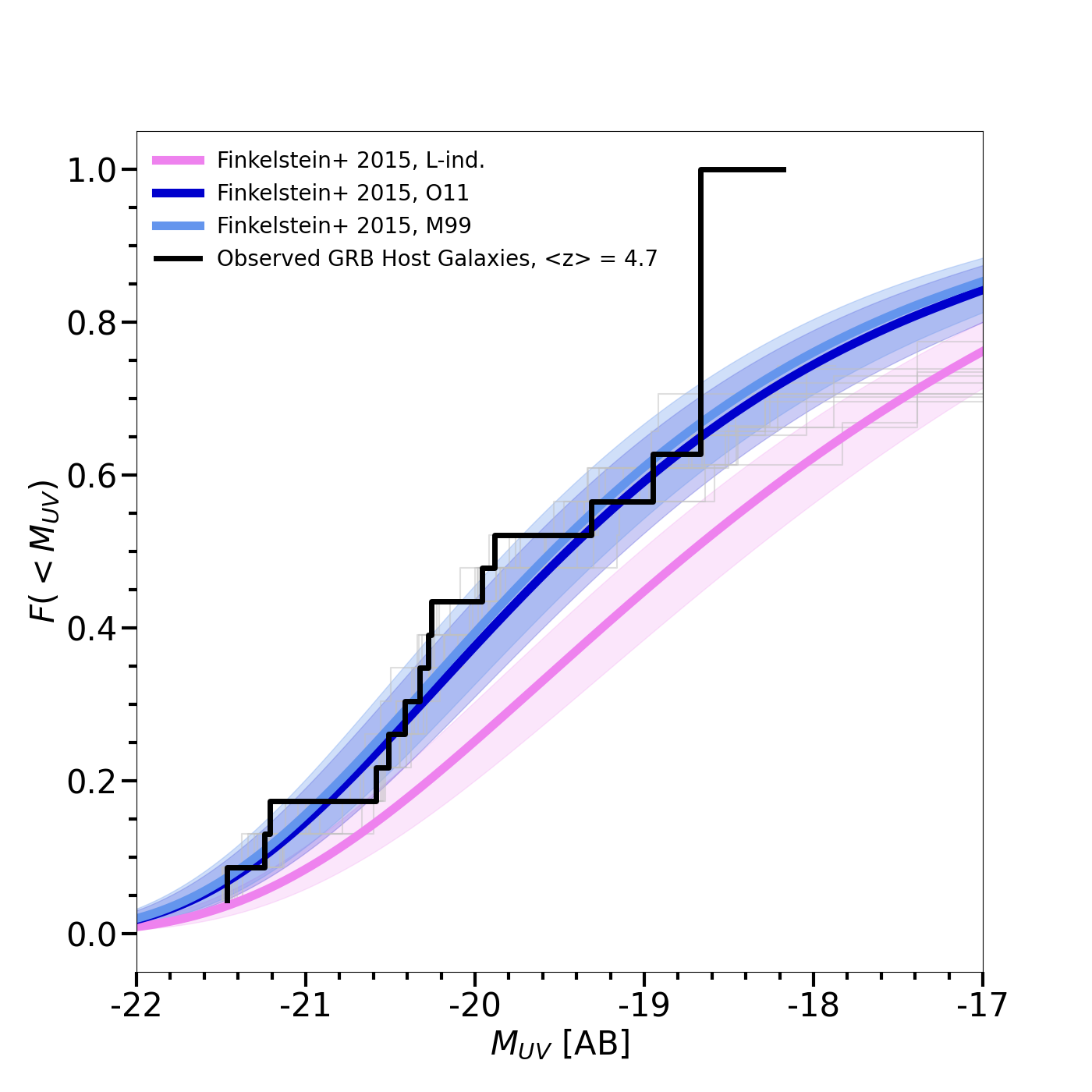}
    
    \caption{\textbf{Left:} Observed cumulative distribution function (CDF) of the UV luminosity function for GRB host galaxies at $z\sim5$ [black] compared to the $z\sim5$ CDFs of the SFR-weighted Lyman-break galaxies from \cite{Bouwens_2021a} with the luminosity-independent galaxy extinction correction in pink and with the empirical luminosity-dependent galaxy extinction from \cite{Bouwens2015} and \cite{Overzier2011} (O11) in dark blue and the empirical luminosity-dependent galaxy extinction from \cite{Bouwens2015} and \cite{Meurer1999} (M99) in light blue. Uncertainties on the Lyman-break galaxy relations are shown as shaded regions and represent 1$\sigma$ uncertainty on the luminosity function parameters [light-pink and light-blues].  Uncertainty on the GRB host galaxy LF is shown with 10 random pulls of the modeled GRB host galaxy magnitudes [silver]. \textbf{Right:} A similar plot to that on the left, with the same GRB host galaxy LF, but using as comparison now the $z\sim5$ SFR-weighted UV LF from \cite{Finkelstein2015}. The pink luminosity function again shows an assumption where galaxy extinction is luminosity independent, while the luminosity function in dark blue (light blue) assumes the O11 (M99) extinction correction.}
    \label{fig:mainLF}
\end{figure*}

We construct a cumulative distribution function (CDF) of the GRB host galaxy luminosity function by using Kaplan-Meier estimation \citep{KaplanMeier1958} on our observed magnitudes and upper limits.  We plot this CDF in Fig. \ref{fig:mainLF}.  We qualify the uncertainty on this CDF by plotting also a subset of the CDFs created from random draws of the modeled magnitude sets.  In this figure we also show the CDFs of the UV LFs from \cite{Bouwens_2021a} and \cite{Finkelstein2015} with the different extinction assumptions.

To measure the likelihood of inconsistency between the Lyman-break galaxy and metallicity-biased GRB host galaxy luminosity distributions to that of the observed $z\sim5$ GRB host galaxy distribution, we use a log-rank test.  This test was chosen because for its applicability to distributions including censored data (i.e., our upper limits), unlike commonly used statistical tests, like a Kolmogorov–Smirnov \citep{kstest} or an Anderson-Darling test \citep{adtest}.  We report the $p$-value corresponding to the calculated $\chi^2$ statistic for each test in Table \ref{tab:LF_pvalues}.  This $p$-value is the probability of achieving the $\chi^2$ test statistic, and so since we consider a $2\sigma$ threshold, we accept $p < 0.05$ as confirmation for the null hypothesis that the compared samples are pulled from different distributions.  To complete these tests, we use \texttt{survdiff} within the \texttt{survival} package in \textit{R} \citep{survival-package,survival-book,R_citation}. 

With $p$-values all above $p = 0.05$, we find no evidence for inconsistency between our O11 and M99 SFR-weighted Lyman-break galaxy luminosity distributions and our derived GRB host galaxy luminosity distribution.  We do, however, find $2\sigma$ disagreement (though $3\sigma$ agreement) between our $L$-independent SFR-weighting for both the B21a and F15 Lyman-break galaxy luminosity distributions and that of our GRB host galaxies.  These results imply that if GRBs are to trace star-formation, either the $L$-independent extinction correction is an incorrect assumption for the distribution of dust in $z\sim5$ star-forming galaxies or additional parameters are necessary, perhaps the faint-end slope curvature parameter $\delta$ presented in \cite{Bouwens2022}.

\begin{figure}
    \centering
    \includegraphics[width=0.5\textwidth]{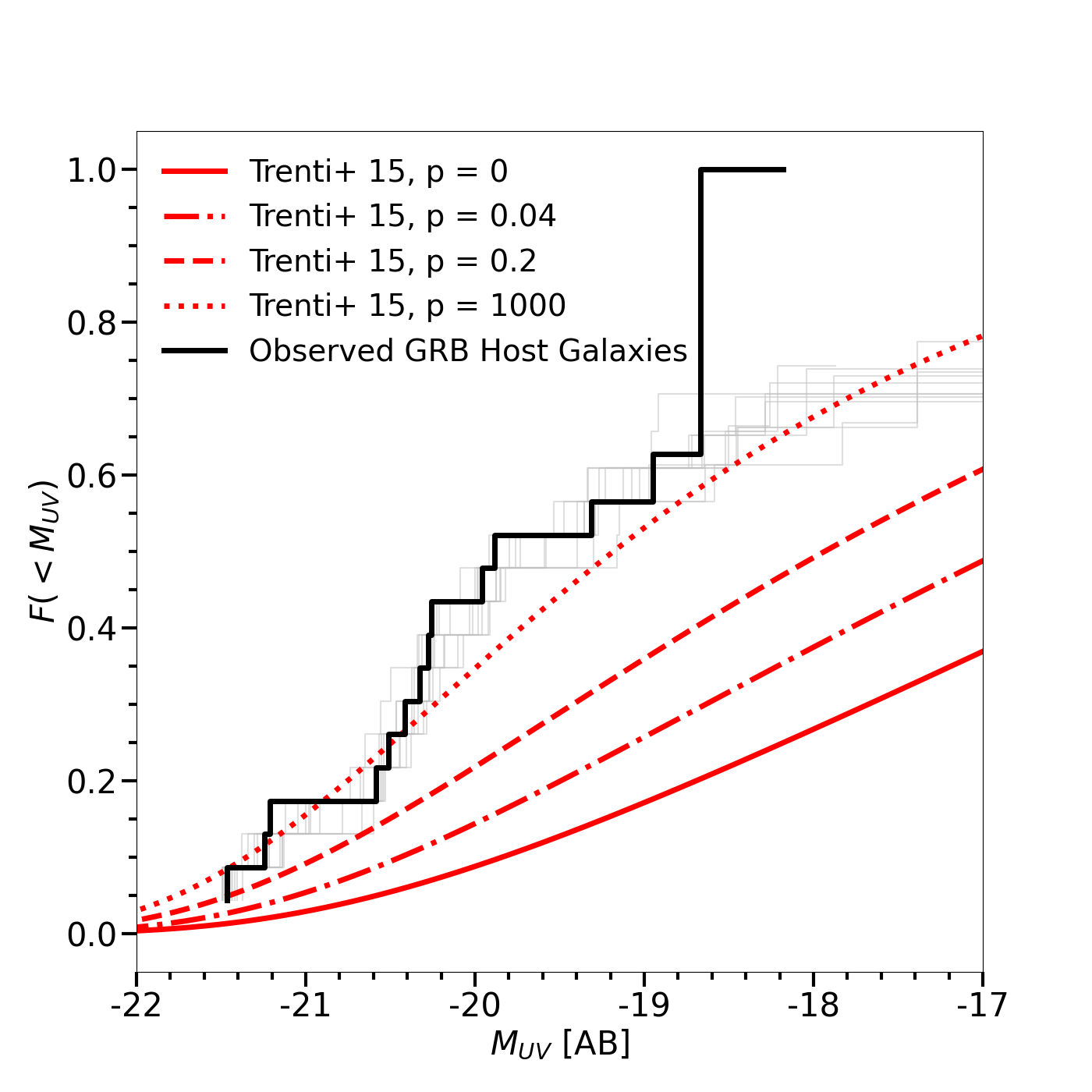}
    
    \caption{Our $z\sim5$ GRB host galaxy LF as described in Figure \ref{fig:mainLF} and predicted GRB host LFs at $z = 4.75$ from \cite{Trenti2015}.  Described in detail in Section \ref{subsec:metallicity}, the \textit{p} parameter is tied to the influence of metallicity on the GRB progenitor path. Across all redshifts, when $p=0$, there is a metallicity bias where GRBs cannot be produced in environments with $Z > Z_{\odot}$, and when $p$ tends to $\infty$, GRB creation is metallicity insensitive.  In \citealt{Trenti2015}, they report Schechter function LF parameters for four choices of $p$, which we plot here.  Results from log-rank tests between the black, median LF for our GRB host sample and the four metallicity-biased LFs are shown in Table  \ref{tab:LF_pvalues}.}
    \label{fig:metLF}
\end{figure}

\begin{deluxetable}{lr}
        \tablecaption{Results of log-rank tests between the $z\sim5$ GRB host galaxy luminosity function derived in this paper and the SFR-weighted Lyman-break galaxy LFs and the predicted metallicity-biased GRB host galaxy LFs. The SFR-weighted Lyman-break galaxy LFs are listed by the different SFR-weights. The $z=4.75$ metallicity-biased LFs are listed by the plateau parameter, as described in Section \ref{subsec:metallicity}.
        \label{tab:LF_pvalues}}
        \tablehead{\colhead{Comparison} &\colhead{log-rank \textit{p}-value}}
        \startdata
            B21a linear & 0.011\\
            B21a O11 & 0.273\\
            B21a M99 & 0.368\\
            F15 linear & 0.007\\
            F15 O11 & 0.167\\
            F15 M99 & 0.228\\
            \hline
            T15 $p=0$ & 2.33e-07\\
            T15 $p=0.04$ & 3.71e-05\\
            T15 $p=0.2$ & 0.002\\
            T15 $p=1000$ & 0.108\\
        \enddata
\end{deluxetable}

\subsection{Investigating the Metallicity Bias}\label{subsec:metallicity}

Lastly, we consider the influence of metallicity in our luminosity function fits.  To quantify GRB-production metallicity sensitivity, we consider the UV LF predictions at $z=4.75$ from \cite{Trenti2015}. Those authors developed a model that considers two GRB progenitor pathways: one that is metallicity-dependent and one that is metallicity independent, which they refer to as  ``metallicity sensitive" (MS) and ``metallicity insensitive" (MI) channels.  They  quantify the percentage of GRBs originating from a MI pathway with their ``GRB efficiency" function, $\kappa(Z)$. This is defined as: $$\kappa(Z) = \kappa_{0} \times \frac{a\log_{10}Z/Z_{\odot} + b + p}{1+p},$$ where $\kappa_{0}, a,$ and $b$ are piecewise defined based on galaxy metallicity and take on the same values as in \cite{Trenti2015}. In this context, $p$ is what they refer to as the ``plateau" parameter and can take on any non-negative value.  

In the low metallicity (and therefore high $z$) limit, this MI efficiency function, $\kappa(Z)$, asymptotically ``plateaus" to the value $p/(1+p)$.  While $p$ is explicitly \textit{not} a probability (and can take on any non-negative value), it is correlated with the percentage of GRBs originating from the MI channel.  Across all metallicities and redshifts, when  $p=0$, it is assumed that GRBs originate exclusively from the MS channel and when $p=\infty$, it is assumed GRBs originate exclusively from the MI channel.  Positive and finite values of $p$ assume a split of GRB progenitor paths. \citealt{Trenti2015} applied their models to the \textit{Swift} GRB catalogue and to other GRB host galaxy samples \citep{Savaglio2009,Cucchiara2015} and found that $p=0.2$ best replicates the redshift evolution of the GRB rate to $z\sim6$.  At $z\sim5,$ the majority of galaxies have metallicities below the threshold values found in the local universe, so we expect the host galaxy LF to be more consistent with the MI parameterization, $p=\infty$.

We show in Fig. \ref{fig:metLF} the four $z=4.75$ luminosity functions predicted by \citealt{Trenti2015} for different values of $p$ overlaid on our GRB host galaxy LF, and we report the results of log-rank tests between these relations in Tab. \ref{tab:LF_pvalues}.  We find only the $p=1000$ case to be consistent with our LF to within the Gaussian-equivalent $2\sigma$.  Specifically, we find disagreement with our observations and the $p=0.2$ model favored by \cite{Trenti2015}. The disagreement of the $p=0.2$ model with the high-redshift host galaxy LFs (ours at $z\sim5$ and that at $z\sim3.5$ from \citealt{Greiner2015}) implies that a different metallicity parameterization for GRB production is necessary.
   
\subsection{GRB Host Size Distribution}
\begin{figure*}
    \centering
    \includegraphics[width=0.8\textwidth]{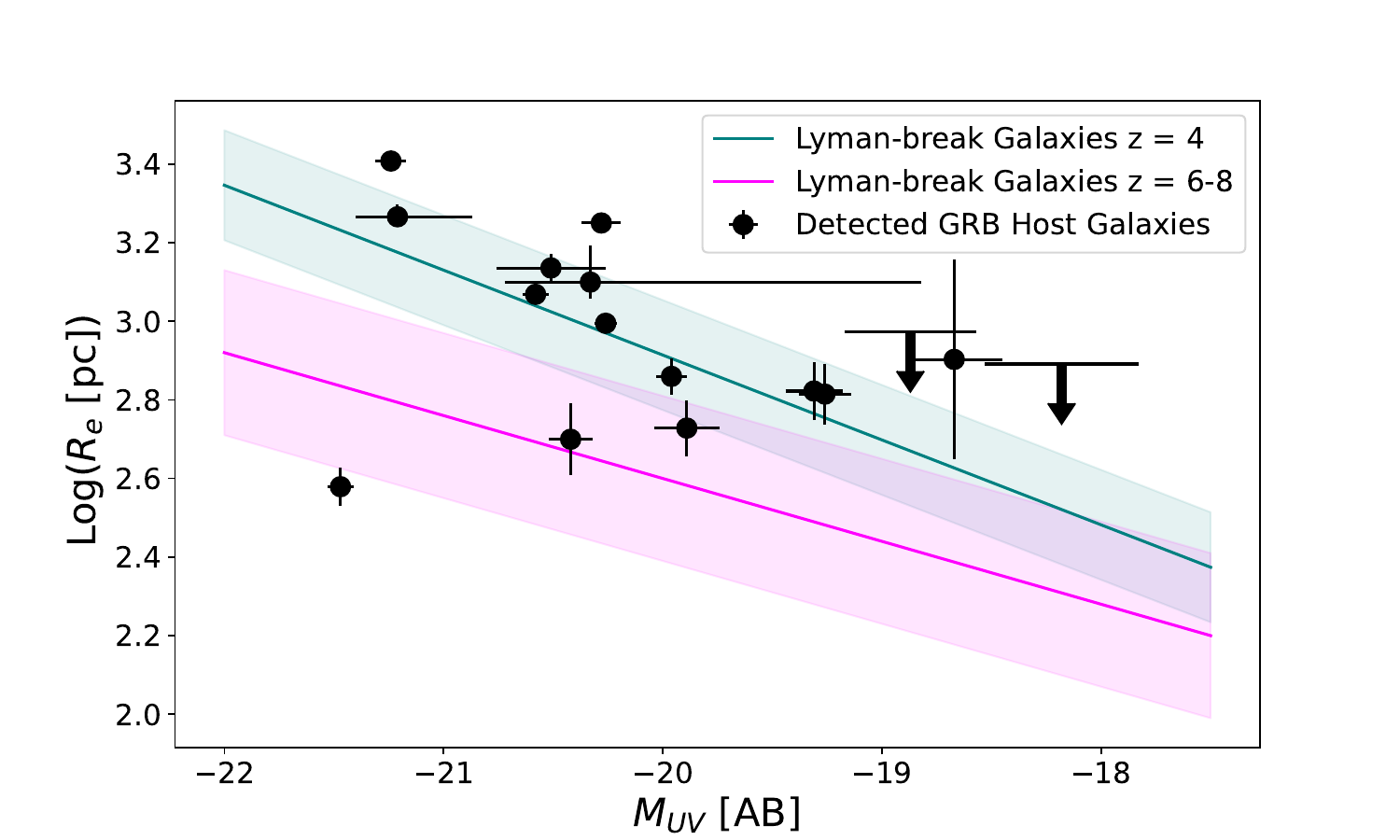} 

    \caption{Galaxy size vs. absolute UV magnitude relation for Lyman-break and GRB host galaxies.  GRB host measurements are shown as black points and are published in Tables \ref{tab:host_info_dets} and \ref{tab:host_sizes}.  The outlier point on the bottom left is that of the host galaxy of GRB 050814. 
 Relations for Lyman-break galaxies at $z = 6-8$ [magenta] and at $z = 4$ [teal] are from \cite{Bouwens_2022_size}.  The shaded regions show the the one sigma scatter of each relation.}
    \label{fig:sizevmuv}
\end{figure*}

Observations of Lyman-break galaxies have shown correlation between the UV luminosity and half-light radius \citep{Kawamata2015,Shibuya2015,Bouwens_2022_size}.  We present half-light radii ($R_{eff}$) for our 16 detected host galaxies in Table \ref{tab:host_sizes} and Figure \ref{fig:sizevmuv}.  We first constructed point spread functions (PSFs) for each of our two filters, F110W and F140W. 
\citet{Schneider2022} found that for a sample of the fields of 42 GRB host galaxies at $z\sim3$ imaged with WFC3/F160W, the constructed PSF had a radius profile stable against the choice of field in which to select stars for the star catalogue but had a S/N dependent on the number of stars selected, increasing with the length of the star catalogue.  In their study of GRB host half-light radii at $z\sim3$, \cite{Schneider2022} find that $N\sim30$ is a sufficient length for the catalogue. We apply this finding to our sample and use 33 stars from the fields of GRBs 050814 and 050922B to construct the PSF for F110W.  The choice of these fields was mostly arbitrary, however we chose not to use fields crowded with several saturated stars (such as that of GRB 140614A).  We had only one GRB field imaged in F140W, and so we select 26 stars from the field of GRB 130606A to construct the PSF for this filter.  We use the \texttt{astropy} package \texttt{EPSFBuilder} \citep{larry_bradley_2023_7946442} to generate the two PSFs from these star catalogues.  

We measure the half-light radii of our detected GRB host galaxies by fitting a S\'ersic light profile with \texttt{GALFIT} \citep{Peng2010}.  On our first measurement attempt, we use as guesses the results from \texttt{Source Extractor} with \texttt{GALFIT} able to fit all parameters.  If the program was not able to converge all parameters, we try again holding $R_{eff}$ constant but all other parameters open.  If the other parameters converge on this run, we fix the parameters to these new values and allow for \texttt{GALFIT} to fit for $R_{eff}$ on the next run.  If the parameters do not converge, or if $R_{eff}$ does not converge as the only free parameter, we instead try fixing all parameters to the \texttt{Source Extractor} guesses and allowing the program to fit for only $R_{eff}$.  If there still was no convergence, and there was a second source within 20 pixels of the host, we rerun the program with a second S\'ersic profile for the second source.  We use the same methods to attempt convergence for both sources.  In all cases, if there was sufficient convergence, the residual was visually checked for confirmation of a good fit.  We record the $R_{eff}$ and its uncertainty reported by \texttt{GALFIT} in Table \ref{tab:host_sizes}. 

No runs were successful following this script, meaning either no convergence of $R_{eff}$ or a visually bad residual, for 3 of our sources (the host galaxies of GRBs 050814, 111008A, and 140311A).  For the host galaxies of GRBs 111008A and 140311A, we adopt $R_{eff}$ upper limits as that reported by \texttt{Source Extractor}.  For the host galaxy of GRB 050814, we updated the \texttt{Source Extractor} guesses to our best guesses, with our only change being updating the position angle (PA) from $-61\deg$ to $40\deg$.  With this update, \texttt{GALFIT} converged all parameters.  This fit is elaborated upon in the Appendix entry for GRB 050814.
 
We compare this sample of GRB host galaxy sizes to Lyman-break galaxy sizes at $z\sim4$ and $z\sim6-8$ \citep{Bouwens_2022_size} in the form of a size-luminosity relation in Figure \ref{fig:sizevmuv}.  Under the assumption that GRBs unbiasedly trace star formation, we expect $z\sim5$ GRB host galaxies to fall in-between the $z\sim4$ and $z\sim7$ relations. Since our smaller sample has a average $z=4.6$, if this assumption is to be true, we would expect the GRB host sample to be weighted closer to the $z\sim4$ relation. We find that $\sim68\%$ (11/16) of our GRB host galaxies fall within or below the 1$\sigma$ scatter of the $z\sim4$ relation. This supports our claim that at $z\sim5,$ Lyman-break and GRB host galaxies trace the same stellar populations.

\subsection{GRB Host Galaxy Non-Detection Fraction and Implications of Hidden Star Formation}
The source of the UV photons needed to reionize the intergalactic medium in the early Universe has been and continues to be an area of very active research \citep{Furlanetto2009,Robertson2015,Endsley2023}.  One explanation for this process is the UV radiation from massive stars in star-forming galaxies \citep{Madau1999, Ciardi2000, Bunker2004,Finkelstein2010, Bunker2010, Finkelstein2012}. Until recently, with the launch of \textit{JWST}, investigations of the feasibility of this explanation have mostly stopped at $z\sim8$ or have relied on the extrapolation of the characterizations of lower redshift observations of Lyman-break galaxies to higher redshifts and fainter magnitudes \citep{Oesch2010, Bouwens2012_reionization}.  Recent \textit{JWST}-based studies have found discrepancies from lower-$z$ expectations and models, namely the detection of more massive, bright galaxies than expected \citep{Harikane2023b,CEERS1}.  There have been many offered explanations for these discrepancies, including stochastic star-formation \citep{Furlanetto2022, Shen2023, Mirocha2023} and high-efficiency star formation \citep{Dekel2023}.

While using GRBs to test the feasibility of massive star reionization of the Universe is not new \citep[e.g.,][]{Tanvir2019}, our complete GRB sample offers the first opportunity to test this feasibility with statistical robustness at a redshift just outside the Epoch of Reionization.  From our non-detection fraction, we can estimate the percentage of star-formation that is occurring in galaxies fainter than our detection limit (i.e., galaxies that are intrinsically faint and galaxies that would otherwise be detected but are dust obscured).  These are galaxies that are inherently often missed in star-forming galaxy samples as they are not directly observable.  Comparing the direct measurement of the percentage of faint star formation to expectations from Lyman-break galaxy LFs is critical, as faint star-forming galaxies are thought to be important contributors of ionizing photons \citep{McLure2010}.

Under the assumption that GRBs unbiasedly trace star formation at this redshift, using binomial statistics, our non-detection fraction of 7/23 is consistent at the $95\%$-confidence level with 13--53\% of star formation occurring in galaxies fainter than our detection limit of $M_{UV} \approx  -18.3$ mag.  This measurement is unique in that it is independent of the functional form of the luminosity function and offers a non-parametric way to test the consistency of an assumed functional form to an observed quantity.  It is shown in Figure \ref{fig:mainLF}, that the percentage of undetectable star formation predicted by the SFR-weighted Schechter function Lyman-break galaxy luminosity functions is $\sim40 \pm 5\%$ and $\sim25 \pm 5\%$ when considering $L$-independent and $L$-dependent (O11 and M99) host extinction, respectively. The lack of disagreement between all of the Lyman-break galaxy predictions and the GRB host galaxy measurement offers support for the hypothesis that star-forming galaxies are large contributors of ionizing photons in the early universe.  

\begin{deluxetable}{lrr}
        \tablecaption{Host Galaxy Size. From left to right the columns are the name of the GRB, the half-light radius ($R_{eff}$) in pixels, and $R_{eff}$ in pc.  When applicable, redshift uncertainty was propagated.
        \label{tab:host_sizes}}
        \tablehead{\colhead{GRB} &  \colhead{$R_{eff}$ (pixel)}  & \colhead{$R_{eff}$ (pc)}}
        \startdata
            050505  &  1.14(0.24) &  $501 \pm 105$ \\
            050803  &  2.87(0.20) &  $1258^{+271}_{-119}$ \\
            050814 & 1.00(0.11) & $379 \pm 42$\\
            050922B  &  4.47(0.21) &  $1844^{+135}_{-103}$ \\
            060206  &  1.78(1.04) &  $800 \pm 467$ \\
            060223  &  1.67(0.18) &  $724 \pm 78$ \\
            060510B  &  2.85(0.07) &  $1171 \pm 29$ \\
            071025  &  3.28(0.24) &  $1366^{+113}_{-114}$ \\
            090516A  &  5.72(0.22) &  $2556 \pm 98$ \\
            100513A  &  1.28(0.21) &  $535 \pm 88$ \\
            111008A & $<2.30$ & $<940$\\
            120712A  &  1.47(0.26) &  $652 \pm 115$ \\
            130606A  &  2.64(0.13) &  $988 \pm 49$ \\
            140311A & $<1.90$ & $<778$\\
            140518A  &  1.58(0.27) &  $664 \pm 114$ \\
            140614A  &  4.03(0.21) &  $1778 \pm 93$ \\
        \enddata
\end{deluxetable}
 
\section{Conclusions}\label{sec:conclusion}

We present new rest-frame UV \textit{HST} imaging of a complete sample of 23 long GRB host galaxies at $z\sim5$.  From our imaging, we measure UV magnitudes and galaxy sizes.  We detect 16 GRB host galaxies and place upper limits on the magnitudes of the remaining 7. Of the 16 detections, we are able to spatially resolve 14 and place upper limits on the sizes of the remaining 2.  Through the construction of a UV luminosity function, we find that our GRB host sample is statistically consistent (log-rank test $p>0.05$) with that of the star-forming galaxy population at the same redshift, when using reasonable corrections for the intrinsic extinction in star-forming galaxies.  When investigating the feasibility of a metallicity-bias model of GRBs from \cite{Trenti2015}, we find that our host sample is inconsistent with this model. Assuming a SFR-weighted Schechter-function formalism and a GRB rate proportional to the dust-corrected UV luminosity, we find parametric agreement between both $\alpha$ and $M_*$ of our best-fits and those from \cite{Bouwens_2021a} and \cite{Finkelstein2015}, again regardless of our choice of galaxy extinction.  We find that 11 of our 16 ($\sim68\%$) host galaxies fall within or below the $1\sigma$ scatter of the luminosity-size relation of $z\sim4$ star-forming galaxies from \cite{Bouwens2014b}. The lack of disagreement between the luminosity-dependent UV LFs and the size-luminosity relations between the Lyman-break and GRB host galaxy samples implies that at $z\sim5$, GRBs are unbiased tracers of star formation.

Under this well-supported assumption that GRBs are unbiased tracers of star-formation at this redshift, we use our non-detection fraction of 7/23 and binomial statistics to estimate that, at $95\%$ confidence, 13--53\% of star formation is undetected in observations of these depths.  In other words, we find that up to $\sim50\%$ (or alternatively, only $\sim10\%$) of star formation could be occurring in galaxies with $M_{UV}>-18.3$ mag.  This measurement is complementary to and unique from similar measurements from Lyman-break galaxy surveys since it is insensitive to the parameterization of the luminosity function.  This solidifies the importance of GRB afterglow and host galaxy observations as a tool for studies of high-$z$ star-formation.  

The sample presented here is the largest and most complete sample of GRB host galaxies at this redshift.  It is unlikely that this sample will be surpassed in statistical sensitivity in the near future, due to our bias-minimizing selection cuts.  One of the selection criteria was a detection cut pre-2015.  Since then, there have been 12 additional \textit{Swift}-detected GRBs with $z>4$.  If all of these sources were to meet our sample criteria and followed our detection distribution, the addition of these 12 sources would improve our sensitivity by $\sim40\%$ (i.e., our uncertainty on the Schechter parameters would be reduced by $\sim40\%$).  While an improvement, this precision is still not better than that from Lyman-break galaxy samples and therefore the inclusion would not result in a significant statistical advance from the analysis performed here. What is needed to improve this analysis is, simply, the detection and follow-up of many more high-redshift GRBs. There are missions, like the Space-based multi-band astronomical Variable Objects Monitor (SVOM; \citealt{SVOM}) and the \textit{Gamow Explorer} \citep{Gamow}, planned expressly for such follow-up. The analysis presented here shows directly how results from such missions can be interdisciplinary, improving not GRB science but our understanding of star-forming galaxies as well.

\section{Acknowledgments}
We thank Paolo D'Avanzo for assistance in accessing afterglow images at TNG, Wen-fai Fong for assistance in the $P_{cc}$ calculations, and Kerry Paterson for assistance in afterglow alignment.  We thank Daniel Brethauer, Lindsay DeMarchi, and Natalie LeBaron for their comments which improved this manuscript. 

The TRex team at UC Berkeley is supported in part by the National Science Foundation under Grant No. AST-2221789 and AST-2224255, and by the HeisingSimons Foundation under grant no. 2021-3248. JS acknowledges support from the Packard Foundation. This research made use of Photutils, an Astropy package for detection and photometry of astronomical sources \citep{larry_bradley_2023_7946442}. This work made use of data supplied by the UK Swift Science Data Centre at the University of Leicester. Support for program GO-15644 was provided by NASA through a grant from the Space Telescope Science Institute, which is operated by the Association of Universities for Research in Astronomy, Inc., under NASA contract NAS 5–26555.  This research has made use of the NASA/IPAC Extragalactic Database (NED), which is funded by the National Aeronautics and Space Administration and operated by the California Institute of Technology. This research has made use of the Keck Observatory Archive (KOA), which is operated by the W. M. Keck Observatory and the NASA Exoplanet Science Institute (NExScI), under contract with the National Aeronautics and Space Administration. This research used the facilities of the Italian Center for Astronomical Archive (IA2) operated by INAF at the Astronomical Observatory of Trieste. Based on observations collected at the European Southern Observatory under ESO programs 077.D-0661(B), 083.A-0644(B), 092.A-0124(A), and 093.A-0069(A).

\vspace{5mm}
\facilities{Gemini (GMOS-N/S), HST (WFC3), Keck (LRIS), Palomar Observatory (P60),  TNG (LRS), REM, Swift (XRT and UVOT), VLT (FORS2, X-Shooter)}
\software{Astrodrizzle \citep{AstroDrizzle}, Astropy \citep{astropy:2013, astropy:2018, astropy:2022}, corner \citep{corner}, photutils \citep{larry_bradley_2023_7946442}, \texttt{Source Extractor} \citep{SourceExtractor}, R \citep{R_citation}, RStan \citep{RStan_citation}, \texttt{survival} \citep{survival-package, survival-book} }

\bibliography{myrefs}
\bibliographystyle{aasjournal}

\appendix

\section{Notes on Individual GRB Host Galaxies}\label{sec:appendix}

Positions and uncertainties for all afterglows are reported in Table \ref{tab:afterglow_info}.  Converted absolute magnitudes, Galactic dust corrections, and uncertainties for host galaxy detections are listed in Table \ref{tab:host_info_dets}.  Converted absolute magnitudes and Galactic dust corrections for host galaxy non-detections are listed in Table \ref{tab:host_info_lims}. $5''$ cutouts of the \textit{HST} images centered on the position of the afterglow are presented in Figure \ref{fig:Stamp_Collage}.

\subsection{050502B}

GRB 050502B has a photometric redshift of $z = 5.2 \pm\ 0.3$ as measured from \textit{R} and \textit{I}-band imaging from TNG of the afterglow \citep{050502B_TNG}.  We identify a source in the stacked \textit{I}-band TNG image from 2005-05-03 within the $1.5"$ \textit{Swift}-XRT error circle. The centroid of the afterglow was measured to an uncertainty of $0.041''$ using \texttt{Source Extractor}, and the astrometric alignment to the \textit{HST} image had a measured RMS uncertainty of $0.043"$ using \texttt{TweakReg}.  These uncertainties are added in quadrature for a total positional uncertainty on the afterglow of $0.060''$.  Within a $3\sigma$ ($0.18''$) radius region in the \textit{HST} image at the position of the afterglow, there is no source detected with \texttt{Source Extractor}.  We find two sources within a $5''$ box centered at the position of the afterglow, and we calculate $P_{cc}$ values for both above $0.6$.  For this reason we consider the host of this GRB to be a non-detection.  Following the prescription in Section \ref{sec:obs}, we estimate a limiting magnitude of $m_{F110W} > 27.55$ mag at 3$\sigma$ above background in a $0.37''$-radius aperture.  

\subsection{050505}

GRB 050505 has a spectroscopic redshift of $z = 4.275$ from Keck/LRIS afterglow spectroscopy analyzed in \cite{050505_Keck_LRIS}.  With Keck/LRIS \textit{I}-band imaging from 2005-05-06, we identify a source within the \textit{Swift}-XRT error region.  We estimate an uncertainty on the centroid of $0.0024''$ and an uncertainty on the astrometric alignment to the \textit{HST} image of $0.050''$.  These uncertainties are added in quadrature for a total afterglow positional uncertainty of $0.050''$.  The afterglow and its $3\sigma$ uncertainty region are coincident with a source in the \textit{HST} image.  We calculate a $P_{cc} = 0.02$ for this source, and we identify it as the host of this GRB.  We report a measured apparent magnitude of $m_{F110W} = 25.95 \pm\ 0.10$ mag.  

\subsection{050803}

GRB 050803 has a photometric host-galaxy redshift of $z = 4.3^{+0.60}_{-2.40}$ as detailed in \cite{SHOALS_1}.  All optical afterglow imaging referenced in the GCNs for this source was for a misidentified source at $z=0.4$, and consequently the afterglow position we report is the enhanced \textit{Swift}-XRT position with its $1.4''$ uncertainty. The MAST-assigned WCS for the WFC3/F110W image was incorrect by several arcseconds, however, we were able to correct this with alignment to a WFC3/F160W image (ID: 12307, PI: Levan) of the same field. This alignment has an uncertainty of $0.094''$.  Within the \textit{Swift}-XRT region in the \textit{HST} image, we detect only one source.  This source is consistent with that reported in \cite{SHOALS_1} used to identify the photometric host-galaxy redshift, and we therefore classify it as the host galaxy of this GRB.  Using \texttt{Source Extractor}, we measure an apparent magnitude of this host galaxy of $m_{F110W} = 26.08 \pm\ 0.11$ mag.  We calculate a $P_{cc} = 0.35$ for this source when using the $90\%$-confidence \textit{Swift}-XRT uncertainty as $R_e$. This percent chance coincidence is well above our $10\%$ threshold.  When estimating the impact of false host-association contamination in our sample, we also consider the possibility that this is a non-detection with a measured limiting magnitude of $m_{F110W} > 27.16$ mag.

\subsection{050814}

GRB 050814 has a photometric afterglow redshift of $z=5.77$ as reported in \cite{Curran2008}.  We detect the afterglow in stacked P60 \textit{i}-band imaging from 2005-08-15.  We report an uncertainty on the centroid of the afterglow of $0.16''$, and an uncertainty on the astrometric alignment to the \textit{HST} image of $0.20''$, for a total positional uncertainty of $0.257''$.  Within a $0.78''$-radius region centered at the afterglow position in the \textit{HST} image, we detect a single source for which we calculate a $P_{cc} = 0.08$. We identify this source as the host galaxy of GRB 050814 and measure an apparent magnitude of this galaxy of $m_{F110W} = 25.46 \pm\ 0.03$ mag.

\texttt{GALFIT} was unable to converge on a single S\'ersic profile, following our standard methods of using the \texttt{Source Extractor} parameter results as input.  We were able to achieve converge after modifying the positional angle (PA) guess from -61 deg to 40 deg, our estimate of the PA of the galaxy.  While all parameters converged and the residual image of this solution passed our visual check, the S\'ersic index, $N$, converged to $N = 9.97 \pm 3.03$, which is much larger than expected.  We also attempted to fit the galaxy with two S\'ersic components and achieved convergence for both profiles, but the residual did not pass our visual check. We chose to complete analysis with the $R_{eff}$ from the single component solution, $R_{eff} = 1.00 \pm 0.11$ pixel.

\subsection{050922B}

GRB 050922B has no afterglow detections reported in the literature but has a photometric host redshift of $z = 4.9^{+0.3}_{-0.6} $ as detailed in \cite{SHOALS_1} from  \textit{i}- and \textit{z}-band GTC/OSIRIS imaging.  We detect three sources within the \textit{Swift}-XRT error circle including the source identified in \cite{SHOALS_1}. We measure an apparent magnitude of this source of $m_{F110W} = 25.37 \pm 0.08$ mag. We calculate $P_{cc} = 0.44$ for this source when using the $90\%$-confidence \textit{Swift}-XRT uncertainty as $R_e$.  This percent chance coincidence is well above our $10\%$ threshold.  When estimating the impact of false host-association contamination in our sample, we also consider the possibility that this is a non-detection with a measured limiting magnitude of $m_{F110W} > 27.85$ mag.

\subsection{060206}

GRB 060206 is located at $z = 4.048$ as reported in \cite{060206_redshift}.  We were unable to use \texttt{TweakReg} to align the P60 \textit{R}-band  imaging of the afterglow from 2006-02-06 \citep{060206_P60} to our \textit{HST} image due to there being only one sufficiently bright source in common between the two images.  We instead align each image separately to the \textit{Gaia} DR2 catalogue.  For this alignment, we consider an uncertainty of approximately one \textit{HST} pixel = $0.065''$.  We detect the afterglow with \texttt{Source Extractor} with a positional uncertainty on the centroid of $0.016''$ for a total positional uncertainty of $0.067''$.  Within a $0.20''$ radius region at the position of the afterglow, we detect a source in our \textit{HST} image.  We calculate a $P_{cc} = 0.02$ for this galaxy, and we therefore identify it as the host galaxy of this GRB.  We measure an apparent magnitude of $m_{F110W} = 27.56 \pm 0.22$ mag.

\subsection{060223}

GRB 060223 has a spectroscopic afterglow redshift of $z = 4.406$ as reported by \cite{060223Keckredshift}.  The only afterglow imaging provided in the literature is \textit{V}-band \textit{Swift}-UVOT imaging from 2006-02-23 \citep{060223_Swift}.  There was only one source (a saturated star) in common between the \textit{HST} and UVOT images, so we were unable to complete image alignment using \texttt{TweakReg}. Since each image was aligned to \textit{Gaia} DR2 upon download from their respective archives, we consider the alignment uncertainty to be within one \textit{HST} pixel = $0.065''$.  We add this in quadrature to the afterglow centroid uncertainty measured with \texttt{Source Extractor} of $0.042''$ to get a total afterglow positional uncertainty of $0.077''$.  The afterglow position and its $3\sigma$ ($0.23''$ ) uncertainty region are coincident with a source in the \textit{HST} image.  We calculate a $P_{cc} = 0.06$ for this source and identify it as the host galaxy of GRB 060223.  We measure an apparent magnitude of this galaxy of $m_{F110W} = 26.63 \pm 0.07$ mag.  For this host galaxy, \cite{Blanchard2016} report a Galactic-extinction corrected magnitude of $m_{F110W} = 26.534 \pm 0.069$ mag, which is consistent with our measurement of $m_{F110W} = 26.53 \pm 0.07$ mag.

\subsection{060510B}

GRB 060510B has a spectroscopic afterglow redshift, $z = 4.941$, as measured in \cite{060510B_GMOS_N_Paper}.  We align \textit{i}-band GMOS-N imaging of the afterglow from 2006-05-10 \citep{060510B_GMOS_N_GCN} to the \textit{HST} image.  We measure a RMS alignment uncertainty of $0.09''$, and we add this in quadrature to the afterglow centroid uncertainty of $0.0062''$ measured with \texttt{Source Extractor} for a total afterglow positional uncertainty of $0.09''$.  The afterglow position and its $3\sigma$ ($0.27''$ ) uncertainty region are coincident with a source in the \textit{HST} image.  We calculate a $P_{cc} = 0.04$ for this source and identify it as the host galaxy of this gamma-ray burst. We measure an apparent magnitude of this source of $m_{F110W} = 26.05 \pm 0.06$.  

\subsection{060522}

GRB 060522 has a spectroscopic afterglow redshift of $z = 5.110$ as reported in \cite{060223Keckredshift}.  We reduced \textit{R}-band TNG imaging of the afterglow from 2006-05-22 and report a $0.028''$ uncertainty on the centroid of the afterglow.  We align this reduced image to the \textit{HST} image and report an uncertainty of $0.05''$ on this astrometric alignment.  We sum these uncertainties in quadrature and report a total positional uncertainty of $0.05''$.  We do not detect a source within a $0.15''$-radius region centered at this afterglow position in the \text{HST} image.  We find three sources within a $5''$ box centered at the position of the afterglow, and we calculate $P_{cc}$ values for all above $0.2$.  For this reason we consider the host of this GRB to be a non-detection.  We report a limiting magnitude of $m_{F110W} > 27.83$ mag.  For this source, \cite{Blanchard2016} report a non-detection and an upper limit of $m_{F110W} > 28.9$ mag, and \cite{Tanvir2012} report a non-detection and an upper limit of $m_{F110W} > 28.13$ mag.  \cite{Blanchard2016} define their 3 sigma upper limits as the magnitude at which sources are detected at 3 sigma significance. The result from \cite{Tanvir2012} is inconsistent with our upper limit, however they perform forced photometry in a $0.4''-radius$ aperture at the afterglow location and also use a $2\sigma$ detection threshold.  When we apply the same methods, we are able to reproduce their limit.  For consistency of our GRB host galaxy sample, we continue with our limit of $m_{F110W} > 27.83$ mag.

\subsection{060927}

GRB 060927 has a spectroscopic afterglow redshift of $z = 5.467$ as detailed in \cite{Ruiz-Velasco2007} from VLT/FORS1 spectroscopy.  We are unable to find the centroid of the afterglow with \texttt{Source Extractor} due to blending with a nearby galaxy in \textit{I}-band VLT imaging at 2.6 days post-trigger \citep{Ruiz-Velasco2007}, but the afterglow is visible in \texttt{DS9} after adjusting the scale and smoothing parameters.  We are able to estimate the center of the afterglow to within $0.5$ VLT pixels ($0.126''$), and we also report a $0.023''$ astrometric uncertainty, resulting in a total positional uncertainty of $0.128''$ for the afterglow.  There are no sources detected within $0.385''$ of this position. Within a $5''$ box centered at the position of the afterglow, we find three sources and calculate $P_{cc}$ values for two of them above $0.8$.  The third source (the only one visible with our scaling choice in Figure \ref{fig:Stamp_Collage} and is the blended source in the VLT imaging) had a $P_{cc} = 0.11$.  This nearby source was detected in VLT \textit{R}-band imaging \citep{Basa2012} and therefore is at $z<4$, and we therefore exclude this source as the possible host galaxy for GRB 060927.  For these reasons, we consider the host of this GRB to be a non-detection.  We report a limiting magnitude of $m_{F110W} > 27.84$ mag. \cite{Tanvir2012} report a limiting magnitude $m_{F110W} > 28.57$, however they perform forced photometry in a $0.4''-radius$ aperture at the afterglow location and also use a $2\sigma$ detection threshold.  When we apply the same methods, we are able to reproduce their limit.  For consistency of our GRB host galaxy sample, we continue with our limit of $m_{F110W} > 27.84$ mag.

\subsection{071025}

GRB 071025 has a photometric afterglow redshift of z = $4.8 \pm 0.4$ as presented in \cite{071025_redshift}.  To identify the host galaxy of this GRB, we use \textit{H}-band REM imaging of the afterglow from 2007-10-25.  We were successful in using \texttt{TweakReg} to align the afterglow and \textit{HST} images, despite there being few sources (many of them saturated stars) in common between the fields.  We report an alignment RMS uncertainty of $0.22''$ and an afterglow centroid uncertainty of $0.14''$ for a total positional uncertainty on the afterglow of $0.26''$.  We detect one source within $0.78''$ of the afterglow position in the \textit{HST} image, though this source has a calculated $P_{cc} =  0.12$.  We identify only one other source within $5''$ of the afterglow position: the bright source in the upper left corner in Figure \ref{fig:Stamp_Collage}.  We measure an apparent magnitude of $m_{F110W} = 23.6573$ mag and a $P_{cc} = 0.24$ for this source.  Because of the bright magnitude and higher $P_{cc}$, we elect to identify the first source as the host galaxy of GRB 071025.
We report an apparent magnitude of $m_{F110W} = 26.06 \pm 0.10$ mag for this galaxy.

\subsection{090516A}

GRB 090516A has a spectroscopic afterglow redshift of $z = 4.111$ as reported in \cite{090516A_redshift}. We identify the afterglow in VLT/FORS2 \textit{R}-band imaging from 2009-05-17 and align this imaging to the \textit{HST} image of the field of the GRB.  We report an astrometric alignment uncertainty of $0.022''$ and a centroid positional uncertainty of $0.0058''$ for a total positional uncertainty on the afterglow of $0.023''.$  This position and its $3\sigma$ uncertainty region is directly on a galaxy in the \textit{HST} image.  We calculate a $P_{cc} = 0.08$ for this source and identity it as the host galaxy of GRB 090516A.  We report an apparent magnitude of $m_{F110W} = 25.04 \pm 0.07$ for this galaxy.  This source was also identified by \cite{Greiner2015} and has a reported $M_{UV} = -20.99 \pm 0.4$ mag, which is consistent with our absolute magnitude of $M_{UV} = -21.24 \pm 0.07$ mag.

\subsection{100219A}

GRB 100219A has a spectroscopic afterglow redshift of $z = 4.667$ as measured in \cite{XSGRB}.  We reduce \textit{r}-band Gemini/GMOS-S images from 2010-02-20 of the afterglow \citep{100219A_GMOS_S}. We align this imaging to the textit{HST} image, and we measure an astrometric alignment uncertainty of $0.017''$ and a centroid uncertainty of $0.032''$ for a total positional uncertainty on the afterglow of $0.036''$.  We detect no sources within $0.12''$ of the afterglow position, and we tentatively classify this as a non-detection for the host galaxy of GRB 100219A. We find two sources within a $5''$ box centered at the position of the afterglow. We calculate $P_{cc}$ values of $0.33$ and $0.09$. The galaxy with a $P_{cc} = 0.09$ is the large galaxy to the North-East of the afterglow region in Figure \ref{fig:Stamp_Collage}.  This source is galaxy BN2010 \footnote{\url{http://simbad.cds.unistra.fr/simbad/sim-id?Ident=\%409106632&Name=\%5bBN2010\%5d\%20J101648.52-123357.5\&submit=submit}} at a redshift of $z = 0.217$ as reported in \cite{100219A_GMOS_S}, and is therefore not the host galaxy of GRB 100219A.  For these reason, we consider the host of this GRB to be a non-detection. We estimate a limiting apparent magnitude of $m_{F110W}>27.58$ mag.  \cite{Thone2013} report a $2\sigma$ detection of a source at the position of the afterglow in GTC/OSIRIS \textit{i}-band imaging.  They report $m_{i'} = 26.7 \pm 0.5$, which \cite{Greiner2015} converts to $M_{UV} = -19.74 \pm 0.5$ mag.  This value is inconsistent with our limiting absolute UV magnitude of $M_{UV} > -18.78$ mag.

\subsection{100513A}

GRB 100513A has a redshift of $z = 4.772$ measured from Gemini/GMOS-N afterglow spectroscopy (\cite{Tanvir2019}, \cite{100513A_GMOS_N}). From \textit{R}-band GMOS-N imaging from 2010-05-13, we detect the afterglow and report a $0.0010''$ uncertainty on the astrometric alignment and a $0.022''$ uncertainty on the centroid for a total positional uncertainty on the afterglow of $0.022''$. This position and its $3\sigma$ uncertainty region are coincident with a source in the \textit{HST} image.  We calculate a $P_{cc} = 0.03$ for this source and identity it as the host galaxy of this gamma-ray burst.  We measure an apparent magnitude of $m_{F110W} = 26.65 \pm 0.15$ mag.

\subsection{111008A}

GRB 111008A has a redshift of $z = 4.9898$ as measured from VLT/X-Shooter afterglow spectroscopy as analyzed in \cite{111008A_VLT_XShooter}.  We reduced \textit{R}-band GMOS-S afterglow imaging from 2011-10-09 and aligned it to the \textit{HST} image. We report an astrometric alignment uncertainty of $0.061''$ and an uncertainty on the centroid of the afterglow of $0.052''$.  We sum these in quadrature for a total positional uncertainty of $0.080''$.  We detect a source partially within the $3\sigma$ uncertainty region within the \textit{HST} image.  We calculate a $P_{cc} = 0.03$ for this source, and we therefore report the first galaxy as the host of this GRB.  We measure an apparent magnitude of $m_{F110W} = 27.69 \pm 0.3$ mag.  While this source is clearly visible with the standard 'zscale' in \texttt{DS9}, we note that we had to lower the \texttt{Source Extractor} detection thresholds from their default values for it to be identified.  This apparent magnitude converts to $M_{UV} = -18.71 \pm 0.30$ mag, and it is consistent with the limiting magnitude of $M_{UV} > -20.88$ mag reported in \cite{Greiner2015} and \cite{111008A_Sparre2014}.

\subsection{120712A}

GRB 120712A has a redshift of $z = 4.1745$ as reported in \cite{120712A_XShooter} from a VLT/XShooter spectrum of the afterglow.  We reduce \textit{R}-band GMOS-S afterglow imaging from 2012-07-12 and align this to our \textit{HST} image for host identification.  We measure an astrometric alignment uncertainty of $0.058''$ and an uncertainty on the afterglow centroid of $0.0064''$.  We therefore report a total positional uncertainty on the afterglow of $0.058''$.  At the location of the afterglow, we detect a source in the \textit{HST} image with a $P_{cc} = 0.06$, and we identify this source as the host galaxy.  We measure an apparent magnitude of $m_{F110W} = 27.06 \pm 0.12$ mag.

\subsection{130606A}

GRB 130606A has a redshift of $z = 5.913$, as reported in \cite{130606A_MMT_redshift} from MMT/Blue Channel afterglow spectroscopy.  We reduce and align \textit{i}-band Gemini/GMOS-N imaging of the afterglow from 2013-06-07 \citep{Chornock2013} to our \textit{HST}/WFC3/F140W image.  We report an astrometric alignment uncertainty of $0.026''$ and an uncertainty on the centroid of the afterglow of $0.002''$.  The position of the afterglow is on a source in the \textit{HST} image, and we calculate a $P_{cc} = 0.02$ for this galaxy.  We therefore identify this source as the host galaxy of GRB 130606A, in agreement with the host identification of \cite{130606A_McGuire}.  We measure an apparent magnitude of $m_{F140W} = 26.79 \pm 0.05$ mag, while \cite{130606A_McGuire} report $m_{F140W} = 26.34^{+0.14}_{-0.16}$ mag for this source from the same \textit{HST} images.  While our values are inconsistent, when we use the aperture size reported in \cite{130606A_McGuire}, we are able to reproduce their findings.  For consistency of our GRB sample, we choose to use our measurement of $m_{F140W} = 26.79 \pm 0.05$ mag for analysis.

\subsection{131117A}

GRB 131117A is located at $z = 4.042$, as measured from  VLT/XShooter spectra from \cite{131117A_xshooter}.  We reduce and stack \textit{R}-band VLT/XShooter imaging from 2013-11-17 in order to attempt host identification.  Due to the poor quality of the afterglow imaging and few sources visible in the frame ($< 10$) including a saturated star and the afterglow, it was not possible to use \texttt{TweakReg} to align the afterglow image to the \textit{HST} image, however upon visual inspection, the images appear to be aligned to within one \textit{HST} pixel ($0.065''$).  We use \texttt{Source Extractor} to measure an uncertainty on the centroid of the afterglow of $0.053''$ for a combined total positional uncertainty on the afterglow of $0.084''$.  We find no sources in the \textit{HST} image within $3\sigma$ of the afterglow position.  We find two sources within a $5''$ of this location, and we calculate $P_{cc}$ values for both above $0.7$.  For this reason we consider the host of this GRB to be a non-detection.  We estimate a limiting magnitude of $m_{F110W} >27.39$ mag at 3-sigma above background in a $0.37''$-radius aperture.

\subsection{140304A}

GRB 140304A has a redshift of $z = 5.283$ as measured from GTC/OSIRIS afterglow spectroscopy in \cite{140304A_redshift}.  The radio afterglow of this GRB was well detected with the VLA \citep{140304A_VLA}, and we report that position and its uncertainty in Table \ref{tab:afterglow_info}.  We do not detect any sources to within $3\sigma$ of this afterglow position in the \textit{HST} image.  We find four sources within a $5''$ box centered at the position of the afterglow, and we calculate $P_{cc}$ values for all above $0.4$.  For these reasons, we consider the host of this GRB to be a non-detection.  We measure a limiting magnitude for this galaxy of $m_{F110W} > 27.49$ mag.

\subsection{140311A}

GRB 140311A has a redshift of $z = 4.954$ as measured from Gemini/GMOS-N spectroscopy of the afterglow \cite{140311A_GMOS_N}. From \textit{i}-band GMOS-N imaging from 2014-03-12 from the same reference, we measure an uncertainty of $0.058''$ on the astrometric alignment to the \textit{HST} image and an uncertainty on the centroid of the afterglow of $0.016''$. This results in a total afterglow positional uncertainty of $0.060''$.  At the location of the afterglow, we detect a source in our \textit{HST} image with $P_{cc} = 0.03$, and we classify it as the host of GRB 140311A.  We measure an apparent magnitudes of $m_{F110W} = 28.38 \pm 0.35$ mag.

\subsection{140428A}

GRB 140428A is located at a redshift $z = 4.68^{+0.52}_{-0.18}$ as measured from afterglow photometry reported in \cite{140428A_redshift}.  We reduced and aligned \textit{I}-band Keck/LRIS imaging from 2014-04-29 \citep{140428A_LRIS} to our \textit{HST} image and report an astrometric tie uncertainty of $0.061''$. We measure an uncertainty on the centroid of the afterglow of $0.026''$ for a total afterglow positional uncertainty of $0.066''$.  We find no sources in the \textit{HST} image within $0.18''$ of the afterglow position.  We find four sources within a $5''$ box centered at the position of the afterglow, and we calculate $P_{cc}$ values for all above $0.2$.  For this reason, we consider the host galaxy of GRB 140428A to be a non-detection, and we report a limiting magnitude of $m_{f110w} > 27.66$ mag.

\subsection{140518A}

GRB 140518A is at a redshift of $z = 4.7055$ as reported in \cite{140418A_redshift} from GMOS-N afterglow spectroscopy.  We align \textit{i}-band GMOS-N imaging from 2014-05-18 \citep{140518A_GMOS_N} to our \textit{HST} image, and we report an astrometric tie uncertainy of $0.047''$. We also report an uncertainty of $0.002''$ on the centroid of the afterglow for a total positional uncertainty on the afterglow position of $0.047''$.  The afterglow position is coincident with a source in the \textit{HST} image, and we calculate a $P_{cc} = 0.05$ for this source.  We therefore classify this galaxy as the host of this GRB.  We report an apparent magnitude of $m_{F110W} = 27.22 \pm 0.13$ mag.  

\subsection{140614A}

GRB 140614A has a redshift of $z = 4.233$ as reported in GCN 16401 \citep{140614A_XShooter} from VLT/X-Shooter afterglow spectroscopy.  We reduce and align \textit{i'}-band VLT/X-Shooter imaging of the afterglow from 2014-06-14 \citep{140614A_XShooter} to our \textit{HST} image and report an uncertainty on this astrometric alignment of $0.35''$.  We measure an uncertainty of $0.045''$ on the centroid of the afterglow with \texttt{Source Extractor} for a total positional uncertainty on the afterglow of $0.349''$.  We detect a source in the \textit{HST} image within the $3\sigma$ uncertainty region.  We calculate a $P_{cc} = 0.21$.  While this is above our threshold of $P_{cc} = 0.1$, we identify this source as the host of this GRB because the source is close to the center of the uncertainty region.  We report an apparent magnitude for this galaxy of $m_{F110W} = 26.14 \pm 0.09$ mag.

\end{document}